\documentclass[aps,preprint,superscriptaddress,longbibliography]{revtex4-1}
\usepackage{amsmath}
\usepackage{bm}
\usepackage{graphicx}
\usepackage{subfigure}
\usepackage{amsmath}
\usepackage{amssymb}
\usepackage{mathrsfs}
\usepackage{braket}
\begin{document}
\title{  Trotter errors  in digital adiabatic quantum simulation of quantum $\mathbb{Z}_2$ lattice gauge theory}

\author{Xiaopeng Cui}
\affiliation{Department of Physics \&  State Key Laboratory of Surface Physics, Fudan University, Shanghai  200433, China}
\author{Yu Shi}
\email{ yushi@fudan.edu.cn}
\affiliation{Department of Physics \&  State Key Laboratory of Surface Physics, Fudan University, Shanghai 200433, China}

\begin{abstract}
    Trotter decomposition is the basis of the  digital quantum simulation. Asymmetric and symmetric decompositions are used in our  GPU demonstration of  the digital adiabatic quantum simulations of  $2+1$ dimensional quantum $\mathbb{Z}_2$ lattice gauge theory. The actual  errors in  Trotter decompositions are investigated as  functions  of the coupling parameter and the  number of Trotter substeps in each step of the   variation of coupling parameter.  The relative error of energy is shown to be closely related to the Trotter error   usually defined defined in terms of the evolution operators.    They are  much smaller than the  order-of-magnitude estimation.   The error in the symmetric decomposition is much smaller than that in the asymmetric decomposition.  The features of the Trotter errors obtained here are useful in  the experimental implementation of digital  quantum simulation and its numerical demonstration.

    \vspace{1cm}

    International Journal of Modern Physics B (2020) 2050292. 
    
    https://doi.org/10.1142/S0217979220502926
\end{abstract}

\maketitle

\section{Introdution}

Lattice gauge theory (LGT) is the approach to  gauge theories based on discretizing the spacetime or space to a lattice.  The simplest LGT is $\mathbb{Z}_2$ LGT, which was first presented as a quantum spin model~\cite{Wegner1971,Kogut1979,sachdev}. It is  important  in particle physics~\cite{Kogut_1975, Wilson_1974,PhysRevD.98.074503}, condensed matter physics~\cite{Kogut1979,sachdev,Levin2005,Wen2005,fradkin_2013}, as well as quantum computing~\cite{Kitaev_2003,Kitaev_2009,Scode_2012}.  With the progress of quantum computing,    quantum simulation of quantum $\mathbb{Z}_2$ LGT becomes a possibility~\cite{Zohar2017,
Bender_2018,Lamm2019,Schweizer_2019,qz2_cui}, including analog and digital approaches.
Analog quantum simulation is based on mapping the theory to a similar Hamiltonian of a simulating system. Digital quantum simulation is based on Trotter decompositions of the evolution operator~\cite{Bender_2018}, including asymmetric decomposition~\cite{Trotter1959,Lloyd1996,qz2_cui} and symmetric  decomposition~\cite{Hatano2005, Childs_2019,qz2_cui}, among others. The computational  complexity, expressed in terms of the  number of steps in the  Trotter decomposition, depends directly on the Trotter error.

Recently, a digital quantum simulation of quantum $\mathbb{Z}_2$ LGT is designed   using  quantum adiabatic algorithm implemented in terms of universal quantum gates, and a classical demonstration of this scheme was made thoroughly  in a GPU simulator~\cite{qz2_cui}. Dubbed pseudoquantum simulation, classical demonstration of quantum simulation in  state-of-art fast computers   facilitates the development of quantum algorithms  and quantum softwares, and is also a new approach of computation~\cite{qz2_cui}.

In   real quantum computing experiments,  in order to complete the quantum process before decoherence,  it is crucial to reduce the number of Trotter  steps  as far as the error is acceptable. So it is  important to precisely investigate the   errors.

In this paper,   we perform the  pseudoquantum simulation of  quantum $\mathbb{Z}_2$ LGT, and study how the errors depend on the  step numbers of decompositions. The   accurate   Trotter errors numerically obtained turn out to be much smaller than the previous order-of-magnitude  estimation.  This provides useful information for  experimental implementation of the quantum simulation  and the parameter selection in pseudoquantum simulation.

\section{Order-of-magnitude estimation of the Trotter errors of adiabatic quantum simulation of quantum $\mathbb{Z}_2$ LGT}

\subsection{quantum $\mathbb{Z}_2$ LGT}

Consider the Hamiltonian of the quantum $\mathbb{Z}_2$ LGT defined on a square lattice~\cite{sachdev,Wegner1971},
\begin{equation}
H = Z + gX,
\end{equation}
with
\begin{equation}
X= -\sum_l { \sigma_l^x },
\end{equation}
\begin{equation}
 Z = \sum_{\square} { Z_\square},
 \end{equation}
\begin{equation}
    Z_\square= -\prod_{l \in \square}{\sigma_l^z},
\end{equation}
where  $g$ is the coupling parameter, $l$ represents links on the square lattice, $\square$ represents a plaquette, the smallest loop formed by links. On a square lattice, a plaquette is  a square.

The adiabatic evolution starts with the  ground state $\ket{\psi_0}$ for $g=0$, in which  $Z_\square = -1$  for each plaquette $\square$~\cite{qz2_cui}. In our original   algorithm,  $g$ is   increased from $0$ adiabatically as $g_{k,m}= (k-1)g_s+m\delta$, $(m=1,\cdots,n)$,  $n$ is the total number of substeps for  each step $k$,   $g_s$ is the increase of $g$ in each step, which lasts time $t_s$,  $\delta=g_s/n$ is the increase of $g$ in each substep $m$.  This generalizes the Trotter asymmetric  and symmetric decompositions to the case that the Hamiltonian varies at each step $m$ of the Trotter decomposition, which is renamed a substep. The  errors in these decompositions were estimated.

Here we  simplify the matter and consider  $g$ vary only at the end of each step, while remain unchanged in the $n$ Trotter substeps within each step, that is,
\begin{equation}
g_{k,m}=g_k= (k-1)g_s, \label{g}
\end{equation}
which remains constant for  $m=1,\cdots,n-1$, and increases only when $m=n$. This is because we shall study the dependence of the error  on $n$.  If $g$ varies at each decomposition step $m$,  the degree of adiabaticity increases with $n$, reducing the error  due to nonadiabaticity~\cite{shi}. To focus on the error due to Trotter decomposition, we now fix the rate of $g$ variation, as given  in (\ref{g}).

\subsection{Definitions and estimation of Trotter errors}

For the asymmetric Trotter decomposition, the error in each step of $g$ variation consisting of $n$ Trotter substeps is
\begin{equation}
\varepsilon^{asy}_s(t_s,n,g)  \equiv \langle \psi(g)| \left( (e^{-iZ\frac{t_s}{n}}e^{-igX\frac{t_s}{n}})^n - e^{-iHt_s}\right) |\psi(g)\rangle.  \label{asymerror}
\end{equation}
For the symmetric Trotter decomposition, the error in each step of $g$ variation consisting of $n$ Trotter substeps is
\begin{equation}
\varepsilon^{sym}_s(t_s, n, g)  \equiv \langle \psi(g)| \left( (e^{-iZ\frac{t_s}{2n}}e^{-igX\frac{t_s}{n}} e^{-iZ\frac{t_s}{2n}})^n- e^{-iHt_s}\right) |\psi(g)\rangle. \label{symerror}
\end{equation}

One can estimate the Trotter errors  under the assumption that  $t_s/n$ is   very small~\cite{qz2_cui}.

By using the  identity $e^{A+B}=e^{A}e^{B}e^{-\frac{1}{2}[A,B]+\cdots}$, we obtain
\begin{equation}
 \varepsilon^{asy}_s(t_s,n,g)    \approx O[\frac{1}{2} g N_p n_l \frac{t_s^2}{n}],
\end{equation}
where $N_p$ is the number of plaquettes,  $n_l$ is the number of links in each plaquette. In the derivation,  it has been  considered that each $Z_\square$ is noncommutative with $n_l$ $\sigma^x$'s. $O$ represents the order of magnitude.

For the symmetric Trotter decomposition, by using the identity  $\ln(e^{A/2}e^{B}e^{A/2})= A+B-([A,[A,B]]+2[B,[A,B]])/24+\cdots
$, we obtain
\begin{equation}
   \varepsilon^{sym}_s(t_s, n, g)   \approx O[( \frac{1}{12} g^2  N_p n_l^2 + \frac{1}{24} g N_l n_p^2) \frac{t_s^3}{n^2}],
\end{equation}
where $N_l$ is the number of links, $n_p$ is the number of plaquettes sharing  each link. In the derivation, it has been considered that each $Z_\square$ is noncommutative with $n_l$ $\sigma^x$'s, hence  $[Z_\square,X]$ is the sum of $n_l$  products of one $\sigma^y$ and $n_l-1$  $\sigma^z$'s. Each  $\sigma^y$ is noncommutative with the $n_p$ $Z_\square$'s of the plaquettes sharing with the link $l$.  On the other hand, each  product of one $\sigma^y$ and $n_l-1$ $\sigma^z$'s is noncommutative with $n_l$ $\sigma^x$'s. Therefore  $[Z,[Z,X]]=O(N_pn_ln_p)=O(N_ln_p^2)$, as $N_p=N_ln_p/n_l$, while  $[X,[Z,X]]=O(N_pn_l^2)$. Another way of reasoning is the following. Each $\sigma^x$ is shared by $n_p$ plaquettes, thus $[Z,\sigma^x_l]$ yields $n_p$  products of one $\sigma^y$ and $n_l-1$ $\sigma^z$'s. Each  product  is   noncommutative with $Z_\square$'s of  the  $n_p$ plaquettes, and with the $n_l$ $\sigma^x$'s on the same plaquette. Consequently,   $[Z,[Z,X]]=O[N_l n_p^2]$,   $[X,[Z,X]]=O[N_ln_pn_l]$. With $N_p=N_l n_p/n_l$, this is the same as above.

Therefore  the accumulated total errors are
\begin{equation}
\varepsilon^{asy}(g) = \sum_{k=1}^{N_s(g)}{\varepsilon^{asy}_s(t_s,n, kg_s)} =   \sum_{k=1}^{g/g_s} O( \frac{1}{2} kg_s N_p n_l \frac{t_s^2}{n})
=  O(\frac{g(g+g_s)}{4g_s} N_p n_l \frac{t_s^2}{n})
\end{equation}
for the asymmetric decomposition, and
\begin{equation}
   \begin{aligned}
\varepsilon^{sym}(g) &= \sum_{k=1}^{N_s(g)}{\varepsilon^{sym}_s(t_s,n, kg_s)} =  \sum_{k=1}^{g/g_s} O([ \frac{1}{12} (kg_s)^2  N_p n_l^2 + \frac{1}{24}kg_s N_l n_p^2] \frac{t_s^3}{n^2} ) \\
&= O[\frac{g \left(g+g_s \right) \left(2 g_s n_l^2 N_p+4 g n_l^2 N_p+3 N_l n_p^2\right)}{144 g_s} \frac{t_s^3}{n^2} ] \\
   \end{aligned}
\end{equation}
for the symmetric decomposition.

\begin{figure}[htb]
   \includegraphics[width=0.4\textwidth]{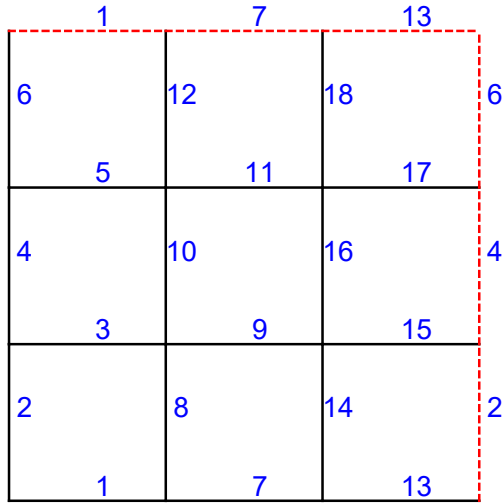}
      \caption{Two-dimensional $3 \times 3$  lattice, with periodic boundary condition. The links are numbered.   }
   \label{fig_lattice}
\end{figure}

In this paper, we consider two-dimensional $3 \times 3$ square lattice  with periodic boundary condition, for which  $N_p=9$, $N_l =18$, $n_p = 2$, $n_l=4$, as shown in Fig.~\ref{fig_lattice}. Therefore
\begin{equation}
\varepsilon^{asy}(g) = O[\frac{9 g \left(g+g_s\right)}{g_s} \frac{t_s^2}{n}],
\end{equation}
\begin{equation}
\varepsilon^{sym}(g) =O[ \frac{g \left(g+g_s\right) \left(288 g_s+576 g+216\right)}{144 g_s} \frac{t_s^3}{n^2}]
\end{equation}

\begin{figure}[htb]
   \centering
      \subfigure[]{}
      \includegraphics[width=0.45\textwidth]{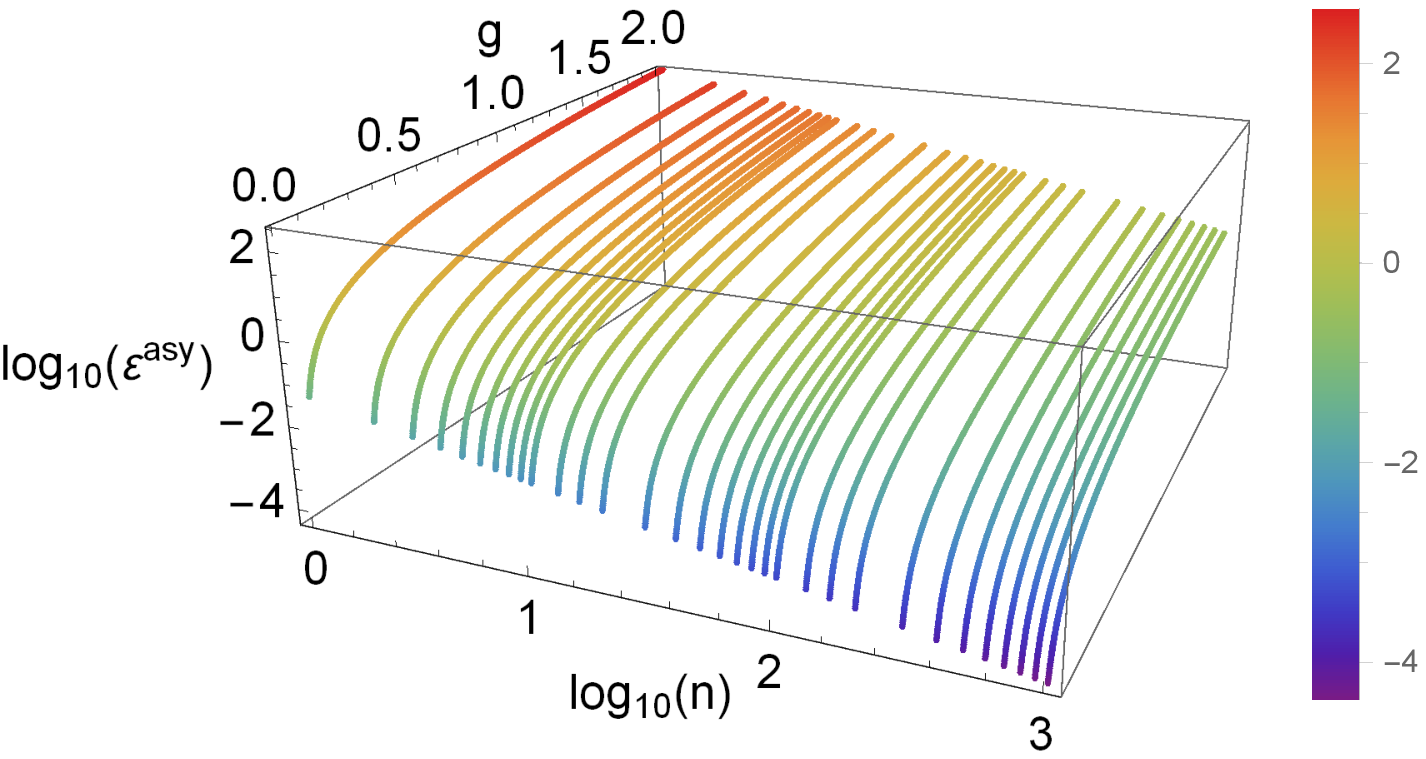}
      \subfigure[]{}
      \includegraphics[width=0.45\textwidth]{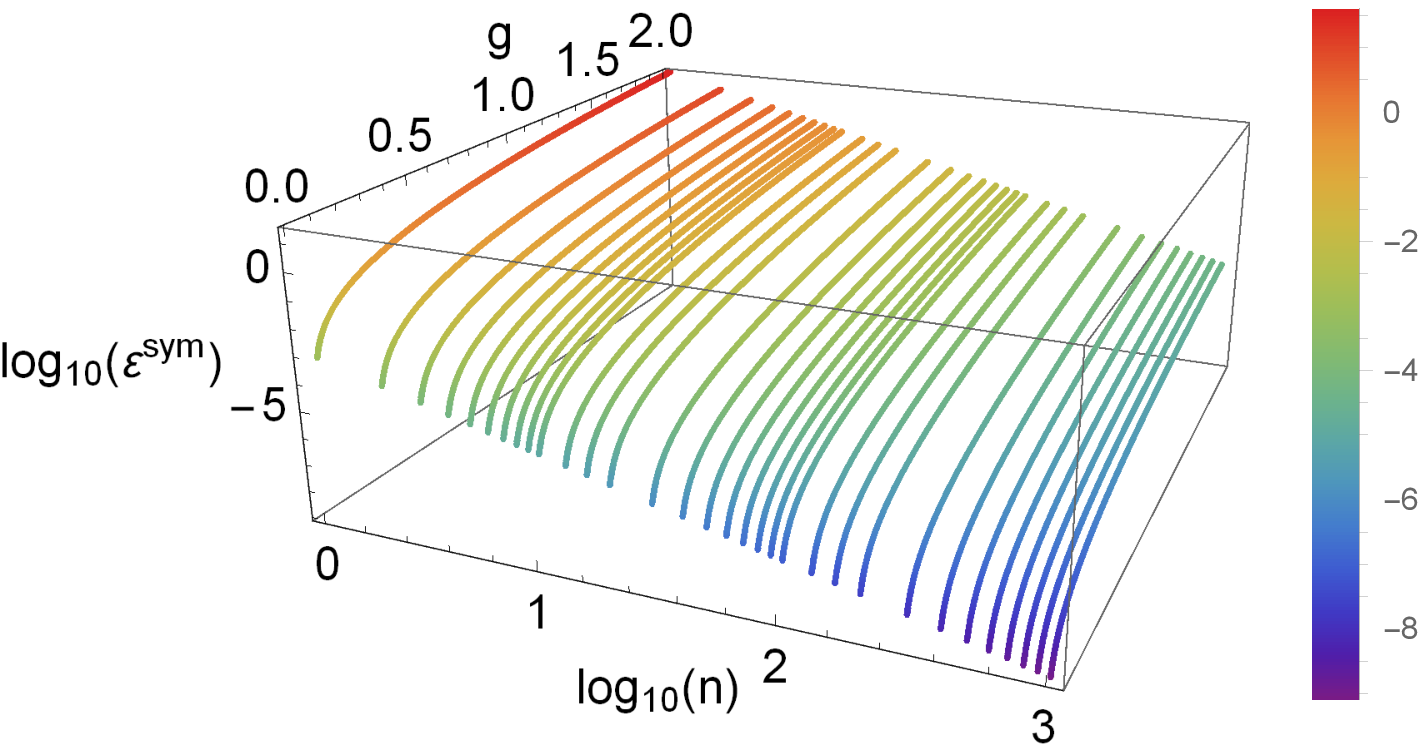}
   \caption{Order-of-magnitude estimated Trotter errors as functions of $g$, which increases from $0$ to $2$ in steps of  $g_s=0.001$ and $t_s=0.1$,  and $\log_{10}n$, which increases from $0$ to $3$.    (a) Accumulated  error  $\varepsilon^{asy}$ for asymmetric Trotter decomposition. (b) Accumulated error  $\varepsilon^{sym}$ for symmetric Trotter decomposition.   }
      \label{fig_estimation}
\end{figure}

Fig.~\ref{fig_estimation} shows the estimation of the errors  as functions of  $g$  and $n$. In our adiabatic simulation, $g$ varies from $0$ to $2$ adiabatically in steps of $g_s=0.001$ and time $t_s=0.1$. It can be seen that the symmetric Trotter has lower error than the asymmetric Trotter. As $t$ is proportional to  $g$ in the adiabatic process, $t= (g/g_s)t_s$, the dependence of the error on $t$ is just the dependence on $(t_s/g_s) g$.

The above order-of-magnitude estimation is  for  very small values of $t_s/n$.  Moreover, it is based on assuming that the order of magnitude of the expectation value of each Pauli matrix is $1$. So in a sense it is an upper bound. Below we calculate the actual error, which could be much lower.

\section{Energy errors and actual  Trotter errors  }

The number of Trotter steps $n$   determines the time  complexity of the adiabatic quantum simulation. Hence it is crucial to know the actual errors. In the following, we describe our method to evaluate the actual errors.

Let's consider the key physical quantity, namely the energy, and define the expectation value
\begin{equation}
E_0(g)\equiv \langle \psi(g)| H(g)|\psi(g) \rangle = {\cal Z}_0(g)+g {\cal X}_0(g),
\end{equation}
with
\begin{equation}
{\cal Z}_0(g)\equiv \langle \psi(g)| Z |\psi(g) \rangle
\end{equation}
\begin{equation}
{\cal X}_0(g)\equiv \langle \psi(g)|X |\psi(g) \rangle\langle  \rangle,
\end{equation}
which are exact,   without   Trotter errors.
Similarly we also consider the expectation values at the state which are calculated using the Trotter decomposition,
\begin{equation}
E_n(g)\equiv \langle H(g)\rangle_n = {\cal Z}_n(g)+ {\cal X}_n(g),
\end{equation}
with
\begin{equation}
{\cal Z}_n(g)\equiv \langle Z\rangle_n,
\end{equation}
\begin{equation}
{\cal X}_n(g)\equiv \langle X \rangle_n,
\end{equation}
where the subscript $n$  means that the number of steps in the Trotter decomposition   is $n$. The difference  $E_n(g)-E_0(g)$ is a  measure  of the error in the Trotter decomposition.

Now we connect energy error  with the Trotter errors $\varepsilon^{asym}_s(t_s, n, g)$ and   $  \varepsilon^{sym}_s(t_s, n, g)$ defined in  (\ref{asymerror}) and (\ref{symerror}), respectively.

Suppose starting from a same  state $|\psi(g)\rangle$, after one step of varying $g$ consisting of $n$ Trotter substep, the exact state is $e^{-iHt_s}|\psi(g)\rangle$, the   state computed by an approximating operator $F$ is  $F|\psi(g)\rangle$. $F=(e^{-iZ\frac{t_s}{n}}e^{-igX\frac{t_s}{n}})^n $ for the  asymmetric Trotter decomposition, and  $F=(e^{-iZ\frac{t_s}{2n}}e^{-igX\frac{t_s}{n}} e^{-iZ\frac{t_s}{2n}})^n$   for the symmetric Trotter decomposition.

Then $E_n(g)-E_0(g) \approx
\langle \psi(g)|  F^\dagger H F |\psi(g)\rangle -
\langle \psi(g)|  e^{-iHt_s} H  e^{-iHt_s} |\psi(g)\rangle$, where we have approximated $|\psi(g-s_s)\rangle$ as  $|\psi(g)\rangle$, because of adiabaticity   $g_s\ll g$. We define the operator
\begin{equation}
R= F- e^{-iHt_s}.
\end{equation}
The expectation value of $R$, $$\varepsilon_s \equiv \langle\psi(g)|R|\psi(g)\rangle,$$  is nothing but the Trotter error as defined in (\ref{asymerror}) and (\ref{symerror}). There, the order-of-magnitude estimation is also made.

$R$  is small,  hence $F^\dagger H F = (e^{iHt_s}+R^\dagger)H (e^{-iHt_s}+R)\approx e^{iHt_s} H  e^{-iHt_s} + e^{iHt_s}HR+ R^\dagger H e^{-iHt_s}$, where the higher order term $R^\dagger R$ is neglected. Therefore, $E_n(g)-E_0(g) \approx \langle \psi(g)|(e^{iHt_s}HR+ R^\dagger H e^{-iHt_s}) |\psi(g)\rangle  $. Using the fact  that $H|\psi(g)=E_0(g)|\psi(g)\rangle$, we have $ \langle\psi(g)|(e^{iE_0(g)t_s} R+e^{-iE_0(g)t_s} R^\dagger)|\psi(g)\rangle= \left(E_n(g)-E_0(g)\right)/E_0(g)$.
We can define
\begin{equation*}
   \begin{aligned}
r(g) &\equiv \langle\psi(g) |(e^{iE_0(g)t_s} R+e^{-iE_0(g)t_s}R^\dagger)|\psi(g)\rangle \\
&=2Re(\varepsilon_s e^{iE_0(g)t_s})\\
&=2|\varepsilon_s|
\cos(E_0(g)t_s+\arg\varepsilon_s), \end{aligned}
\end{equation*}
thus
\begin{equation}
r(g)=\frac{E_n(g)-E_0(g)}{E_0(g)}.
\end{equation}
Therefore we have shown that the Trotter error $\varepsilon_s$, as defined in  (\ref{asymerror}) or  (\ref{symerror}) is simply related to the relative error of the energy.

Thought not directly related to   Trotter errors in  (\ref{asymerror}) or  (\ref{symerror}), the relative errors of $Z$ and $X$ can also be defined similarly as
$$\frac{{\cal Z}_n(g)-{\cal Z}_0(g)}{{\cal Z}_0(g)}, \,\, \frac{{\cal X}_n(g)-{\cal X}_0(g)}{{\cal X}_0(g)}.$$

\begin{figure}[htb]
   \includegraphics[width=0.5\textwidth]{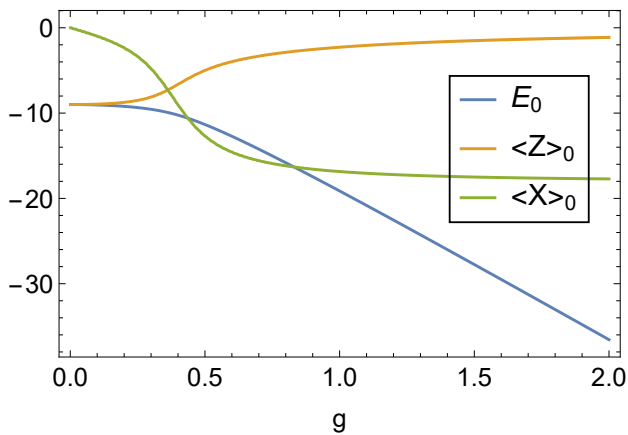}
   \caption{Good approximations of the expectation values  $E_0$, ${\cal Z}_0\equiv \langle Z\rangle_0$ and ${\cal X}_0\equiv \langle X\rangle_0$ of $H$, $Z$ and $X$, respectively, without    Trotter error, as functions  of $g$, obtained by using the symmetric Trotter decomposition  with steps of  $g_s=0.001$, $ t_s=0.1$ and  $n=10^4$.  }
   \label{fig_QPT}
\end{figure}

There are no exact solutions of ${\cal Z}_0(g)$,  ${\cal X}_0(g)$ and $E_0(g)$. Nevertheless,  we can obtain  very good approximations  by  using  symmetric Trotter decomposition with very large number of Trotter steps, say,  $n=10^4$, and with the same values of  $g_s=0.001$ and $t_s=0.1$ as for $E_n(g)$. This   benchmark has the additional advantage that there is some cancellation of the nonadiabatic errors, even though they are small, which exist as the variation of $g$ is not infinitesimally slow. $E_0(g)$, ${\cal Z}_0(g)$ and ${\cal X}_0(g)$ obtained in this way are  shown in Fig.~\ref{fig_QPT}.

\section{Numerical results in the pseudoquantum simulation }

We perform pseudoquantum simulation using  the QuEST GPU quantum simulator with double precision~\cite{quest}, and using Tesla V100 card of Nvidia GPU. In our pseudoquantum simulation, for each set of values of $n$ and $g$, we  calculate the expectation values We vary $n$   from $1$ to $1000$, and numerically calculate  the characteristic quantities.

\subsection{ Energy Errors    }

\begin{figure}[htb]
   \centering
   \subfigure[]{}
   \includegraphics[width=0.45\textwidth]{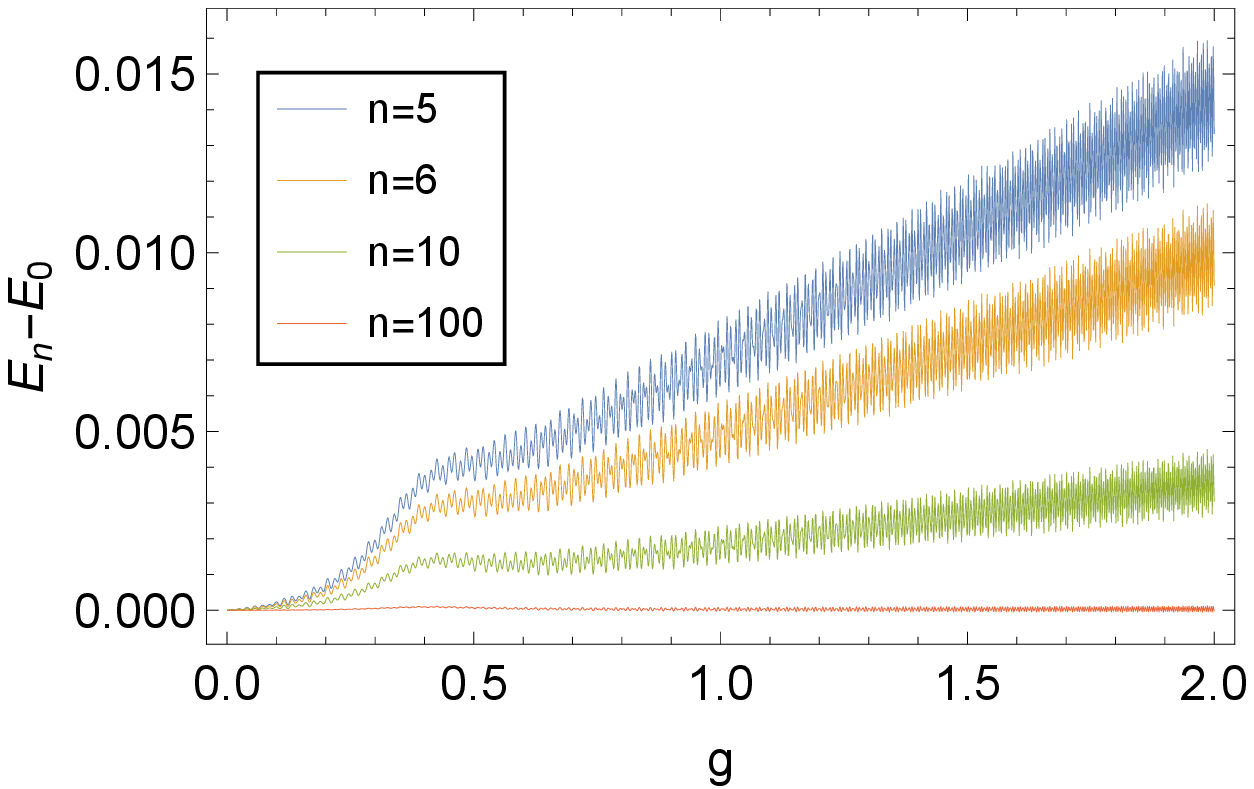}
   \subfigure[]{}
   \includegraphics[width=0.469\textwidth]{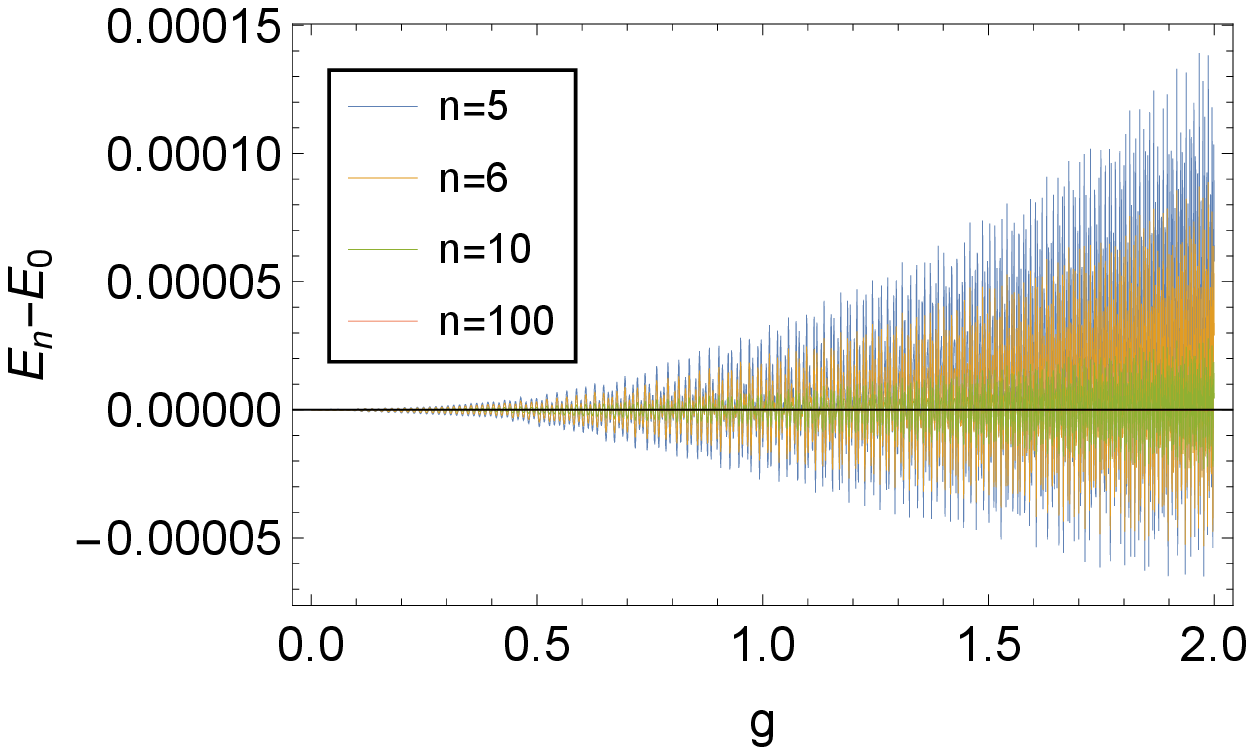}
\subfigure[]{}
   \includegraphics[width=0.45\textwidth]{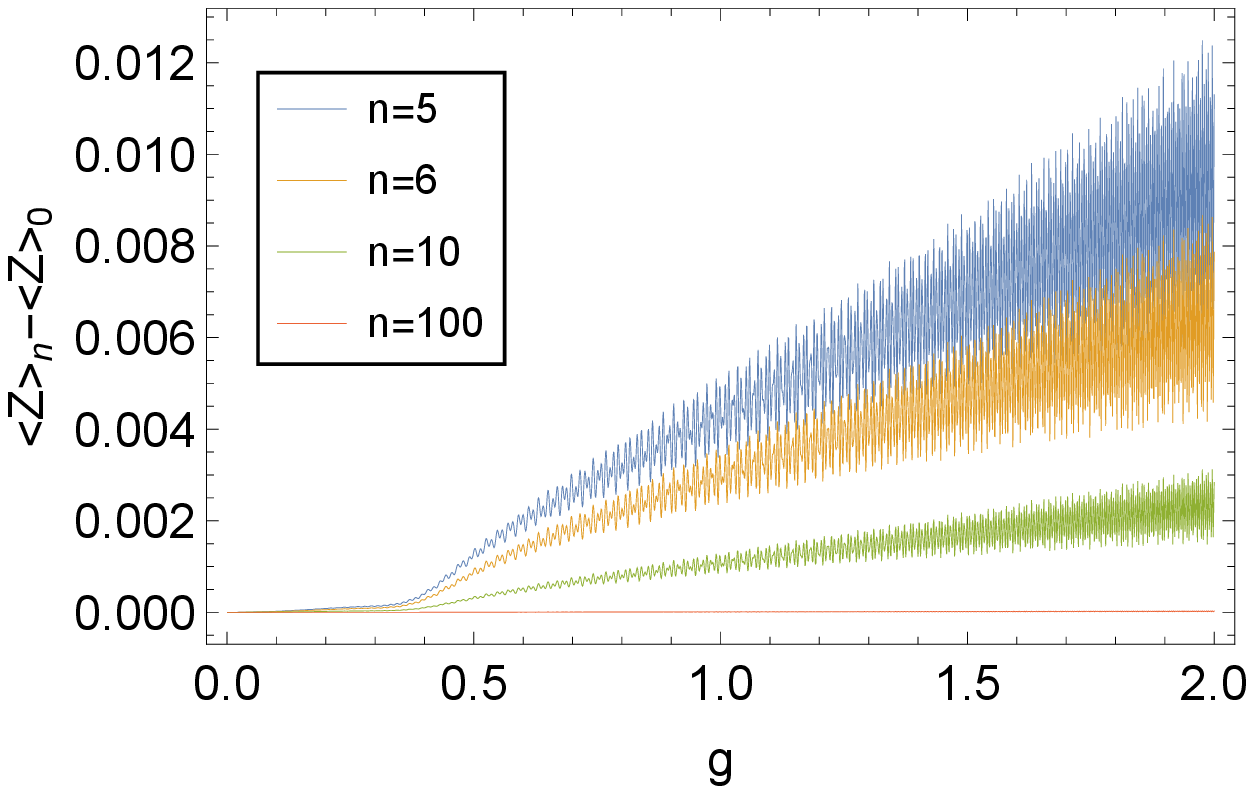}
   \subfigure[]{}
   \includegraphics[width=0.45\textwidth]{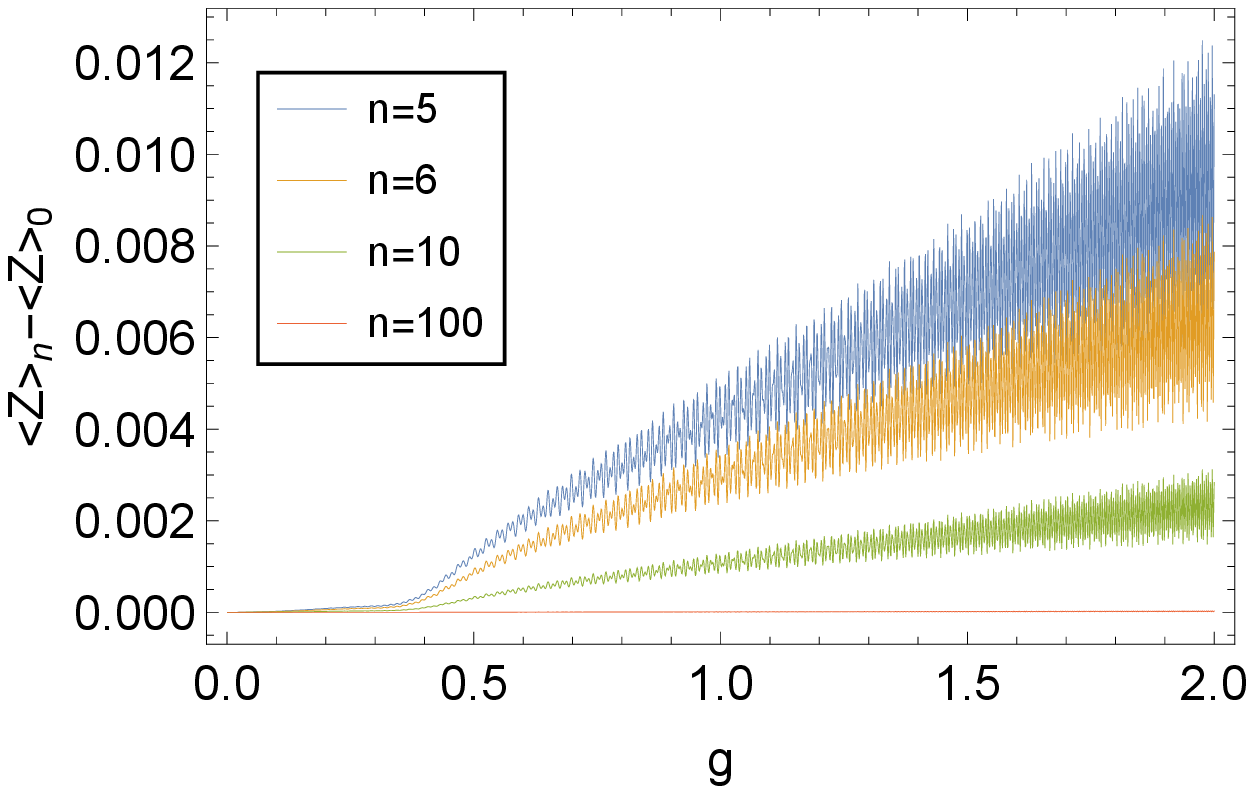}
   \subfigure[]{}
   \includegraphics[width=0.45\textwidth]{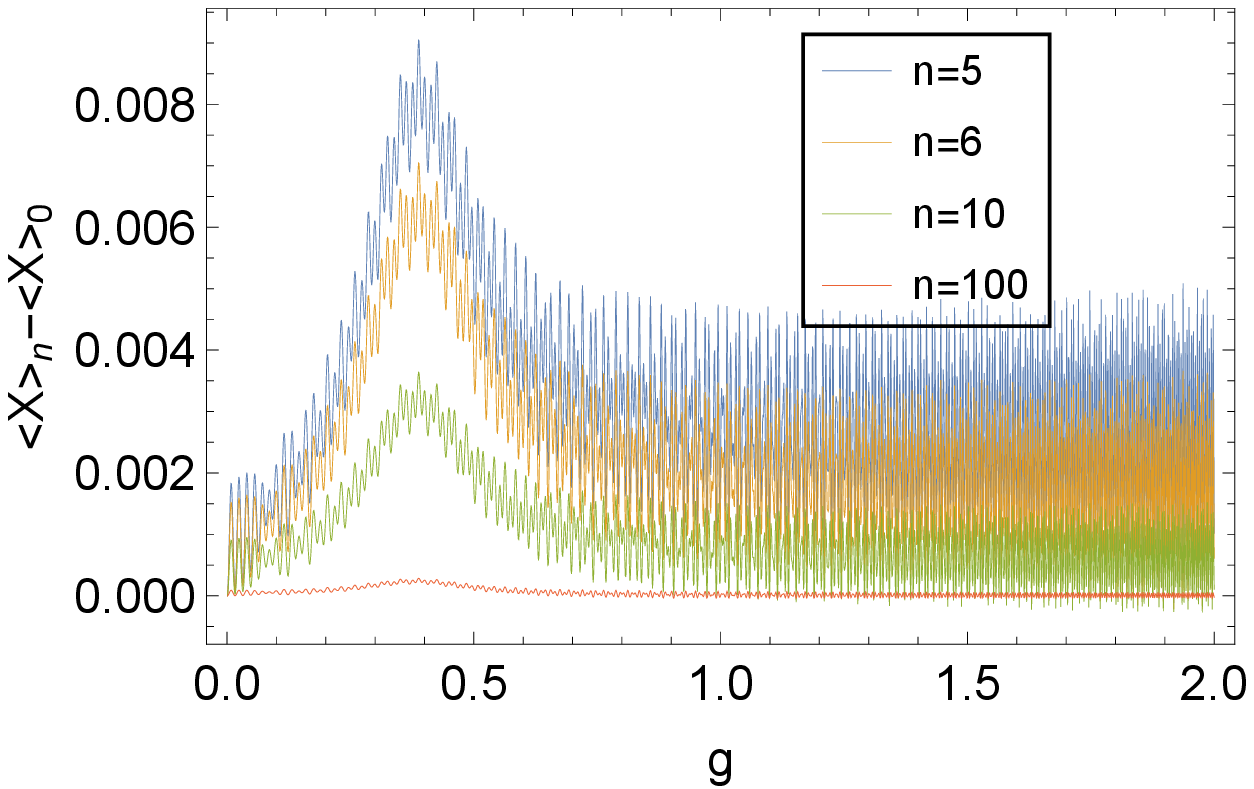}
   \subfigure[]{}
   \includegraphics[width=0.45\textwidth]{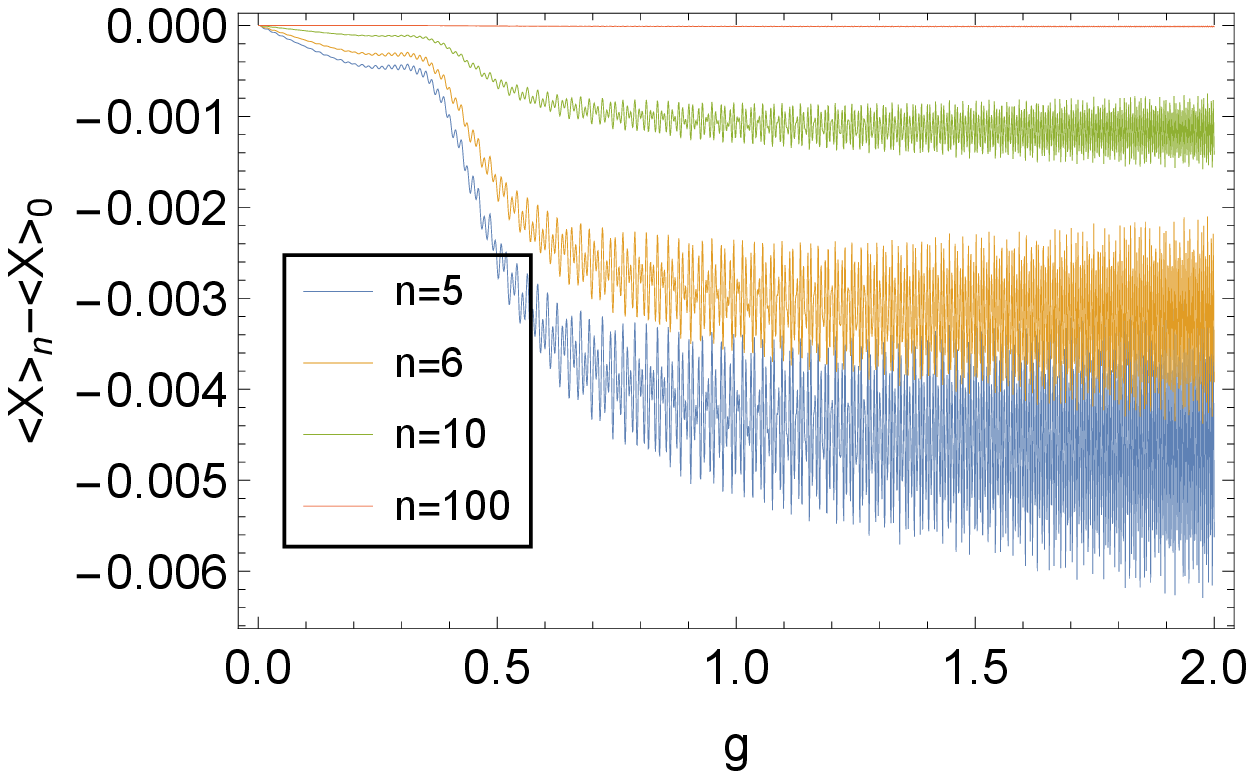}         \caption{Numerical results of errors  as function of $g$, in steps of $g_s =0.001$ with  $n= 5, 6, 10, 100 $.  (a) $E_n-E_0$ in the asymmetric Trotter decomposition. (b) $E_n-E_0$ in the symmetric Trotter decomposition.  (c) $\braket{Z}_n-\braket{Z}_0$ in the  asymmetric Trotter decomposition. (d) $\braket{Z}_n-\braket{Z}_0$ in the symmetric Trotter decomposition. (e) $\braket{\cal X}_n-\braket{X}_0$ in the asymmetric Trotter decomposition. (f) $\braket{\cal X}_n-\braket{X}_0$ in the symmetric Trotter decomposition.}
   \label{fig_E}
\end{figure}

$E_n(g)-E_0(g)$ is shown in Fig.~\ref{fig_E}, where it can be seen  that it is two orders of magnitude lower  in the symmetric decomposition  than  in the  asymmetric decomposition.

Fig.~\ref{fig_E} also shows
${\cal Z}_n(g)-{\cal Z}_0(g)$ and ${\cal X}_n(g)-{\cal X}_0(g)$, which  are of   the same order of magnitude. Moreover,  ${\cal Z}_n(g)-{\cal Z}_0(g)$ in both decompositions are  nearly the same.  ${\cal Z}_n(g)-{\cal Z}_0(g)$ is positive. In asymmetric decomposition, ${\cal X}_n(g)-{\cal X}_0(g)$ is also positive, hence $E_n(g)-E_0(g)$ is the sum of two positive numbers. In  symmetric decomposition,  ${\cal X}_n(g)-{\cal X}_0(g)$ is    negative, hence $E_n(g)-E_0(g)$ is a sum of one positive number and one negative number. Consequently,  $E_n(g)-E_0(g)$   is significantly smaller in the symmetric decomposition than in asymmetric decomposition.

\subsection{ Error bounds  }

\begin{figure}[htb]
   \centering
   \subfigure[]{}
   \includegraphics[width=0.45\textwidth]{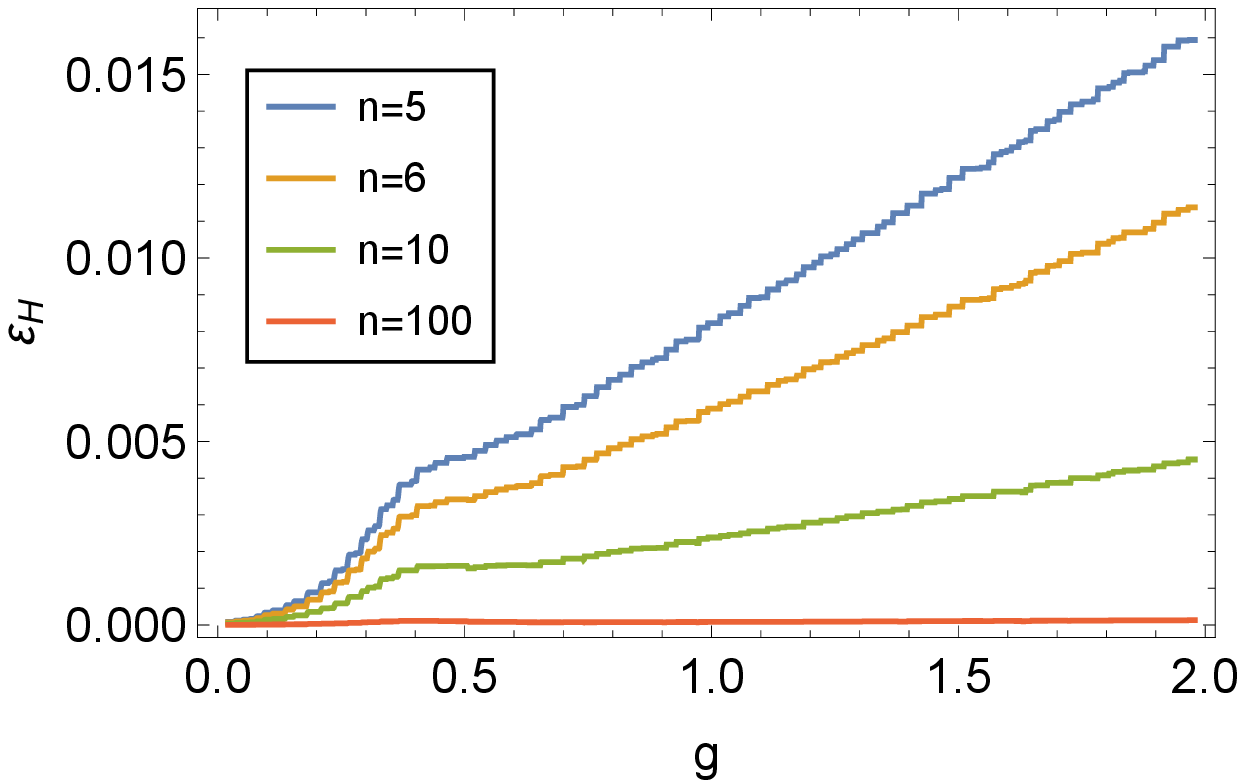}
   \subfigure[]{}
   \includegraphics[width=0.469\textwidth]{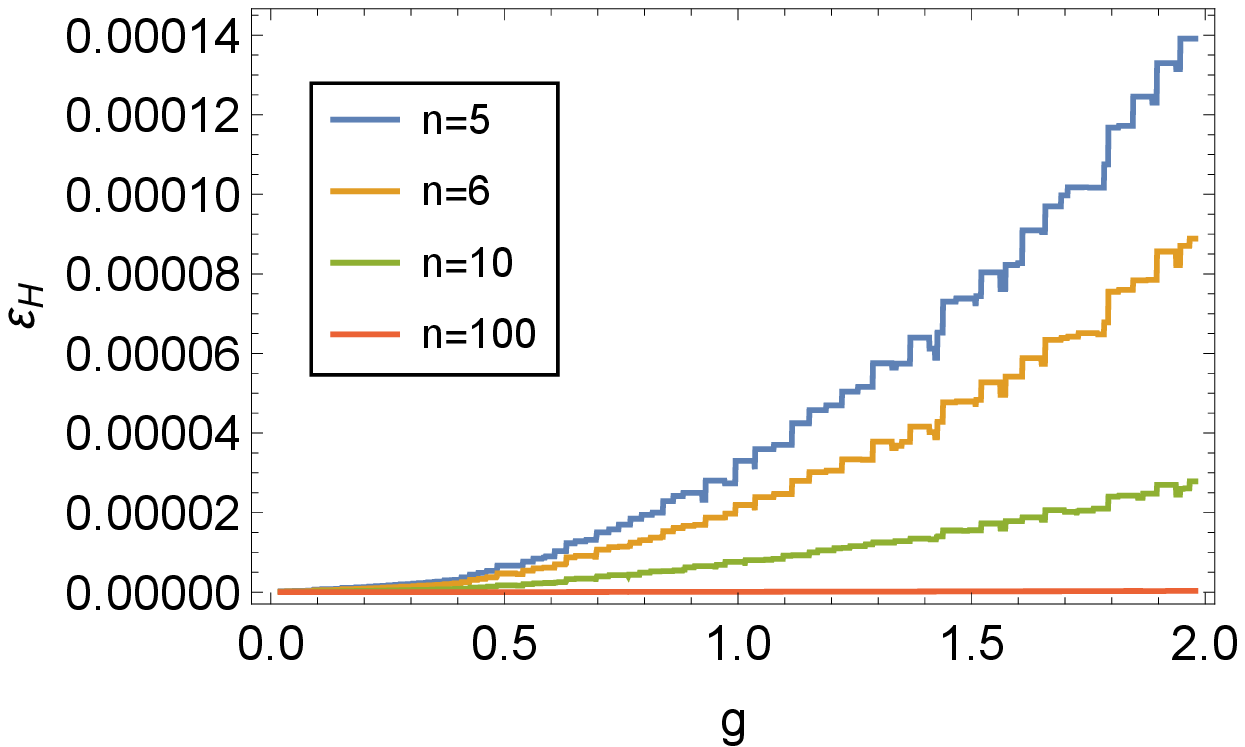}
    \caption{ Error bound  $\varepsilon_H$  calculated from numerical results of $E_n-E_0$,  as a function of g, in step of $g_s =0.001$ with   $n=5, 6, 10, 100$. (a) Asymmetric Trotter decomposition. (b) Symmetric Trotter decomposition.   }\label{fig_smooth}
\end{figure}

\begin{figure}[htb]
   \centering
      \subfigure[]{}
      \includegraphics[width=0.46\textwidth]{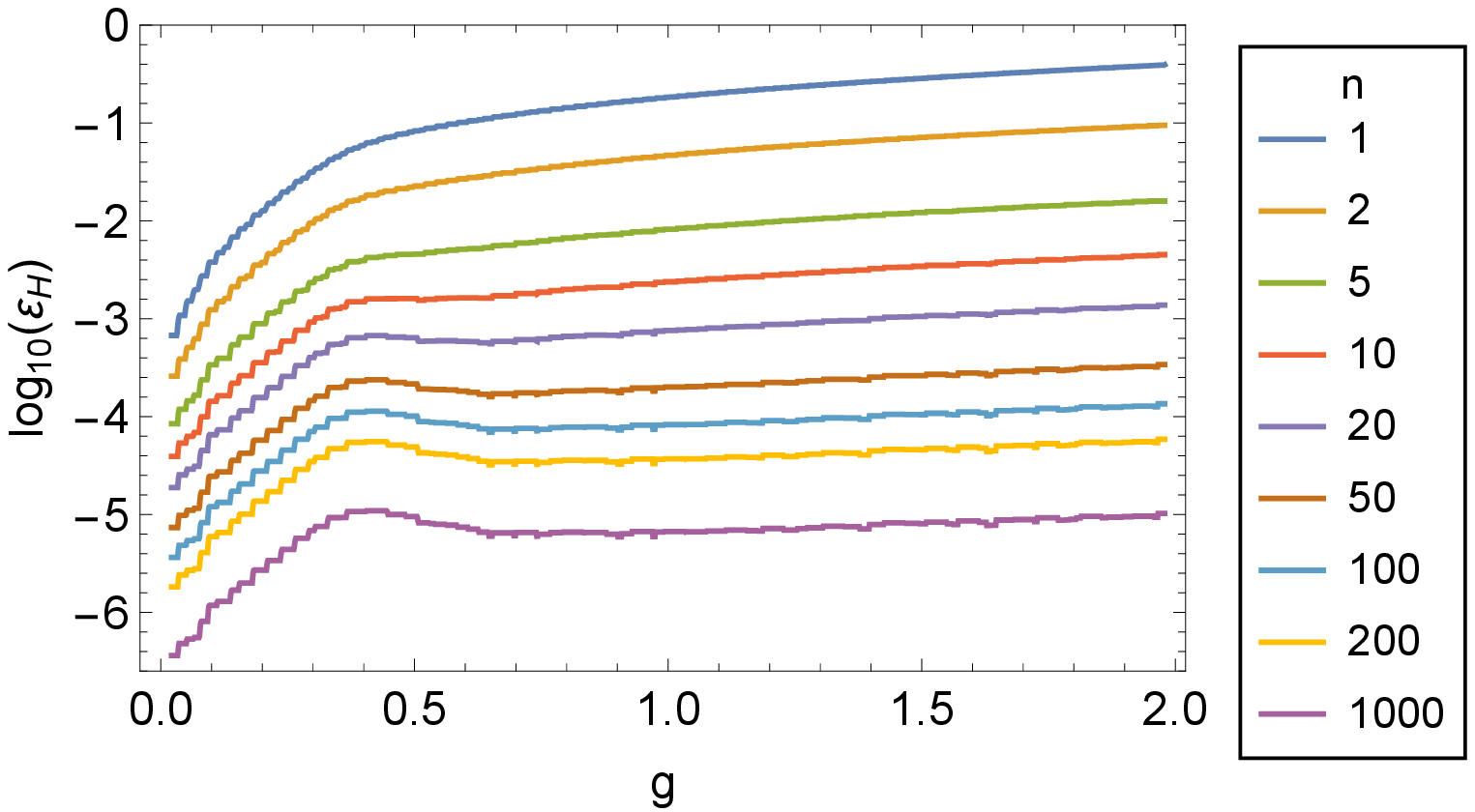}
      \subfigure[]{}
      \includegraphics[width=0.46\textwidth]{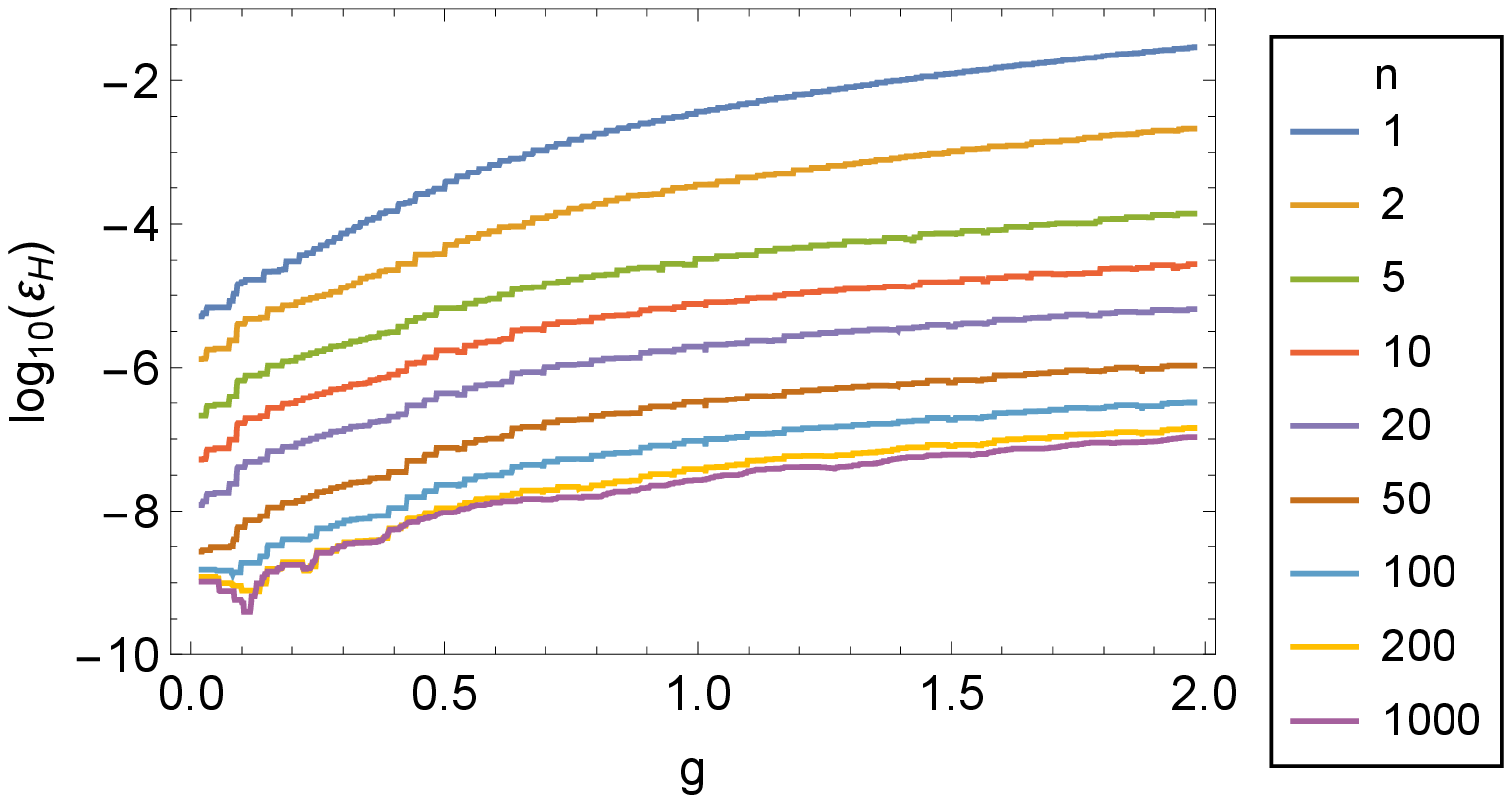}
      \caption{Numerical results of $log_{10}(\varepsilon_H)$ as functions of $g$, in steps of $g_s =0.001$ with various  values of  $n$. (a) Asymmetric Trotter decomposition. (b) Symmetric Trotter decomposition.}
      \label{fig_EB_n}
   \end{figure}

\begin{figure}[htb]
   \centering
      \subfigure[]{}
      \includegraphics[width=0.45\textwidth]{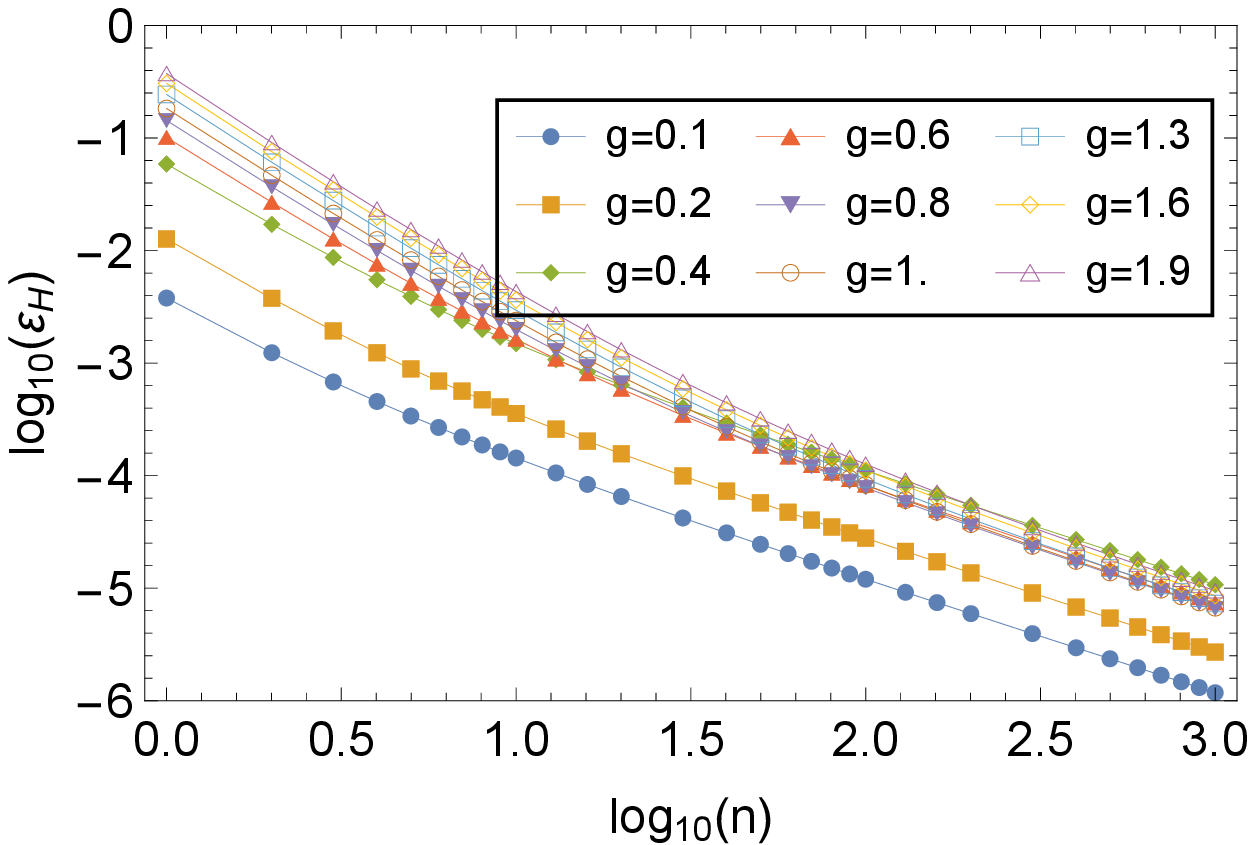}
      \subfigure[]{}
      \includegraphics[width=0.45\textwidth]{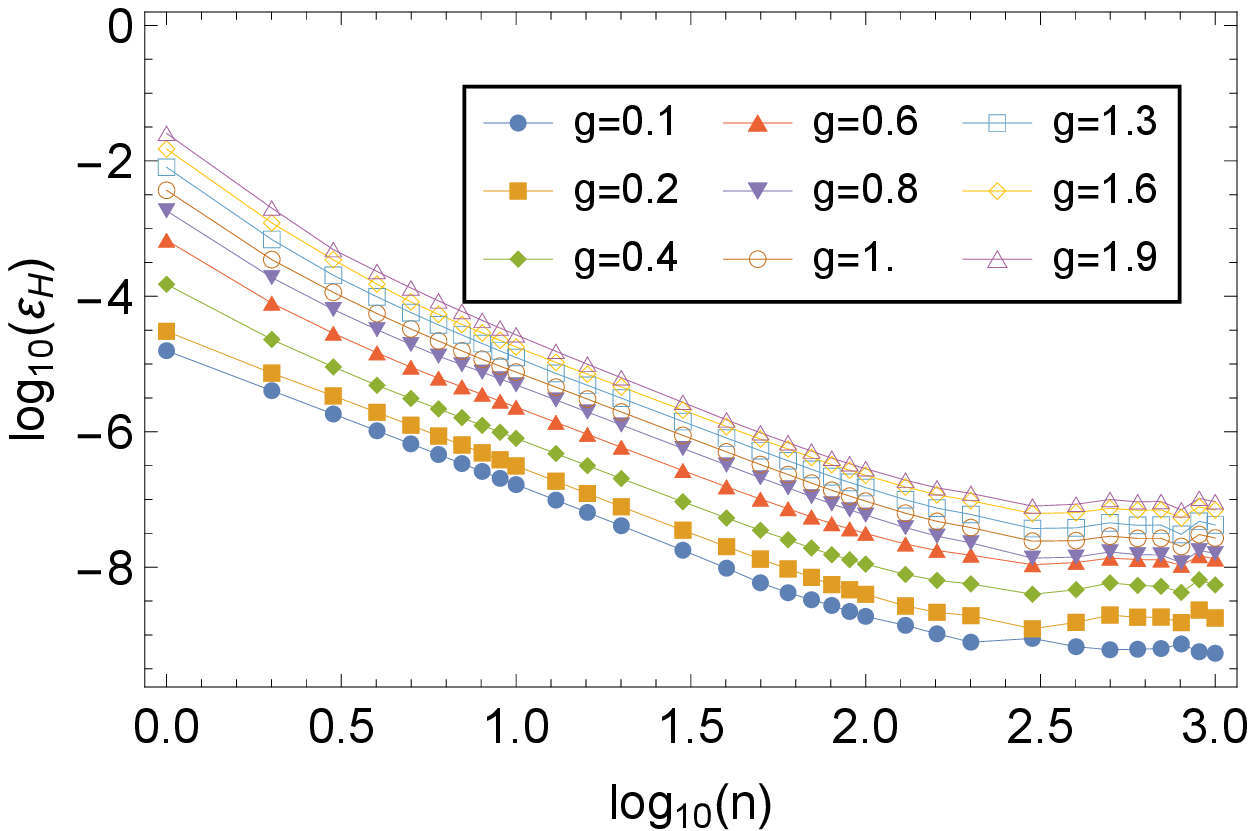}

      \caption{Numerical results of $log_{10}(\varepsilon_H)$ as functions of $n$ for various values of $g$. (a) Asymmetric Trotter decomposition. (b) Symmetric Trotter decomposition.
      }
      \label{fig_EB_g}
   \end{figure}

\begin{figure}[htb]
   \centering
      \subfigure[]{}
      \includegraphics[width=0.45\textwidth]{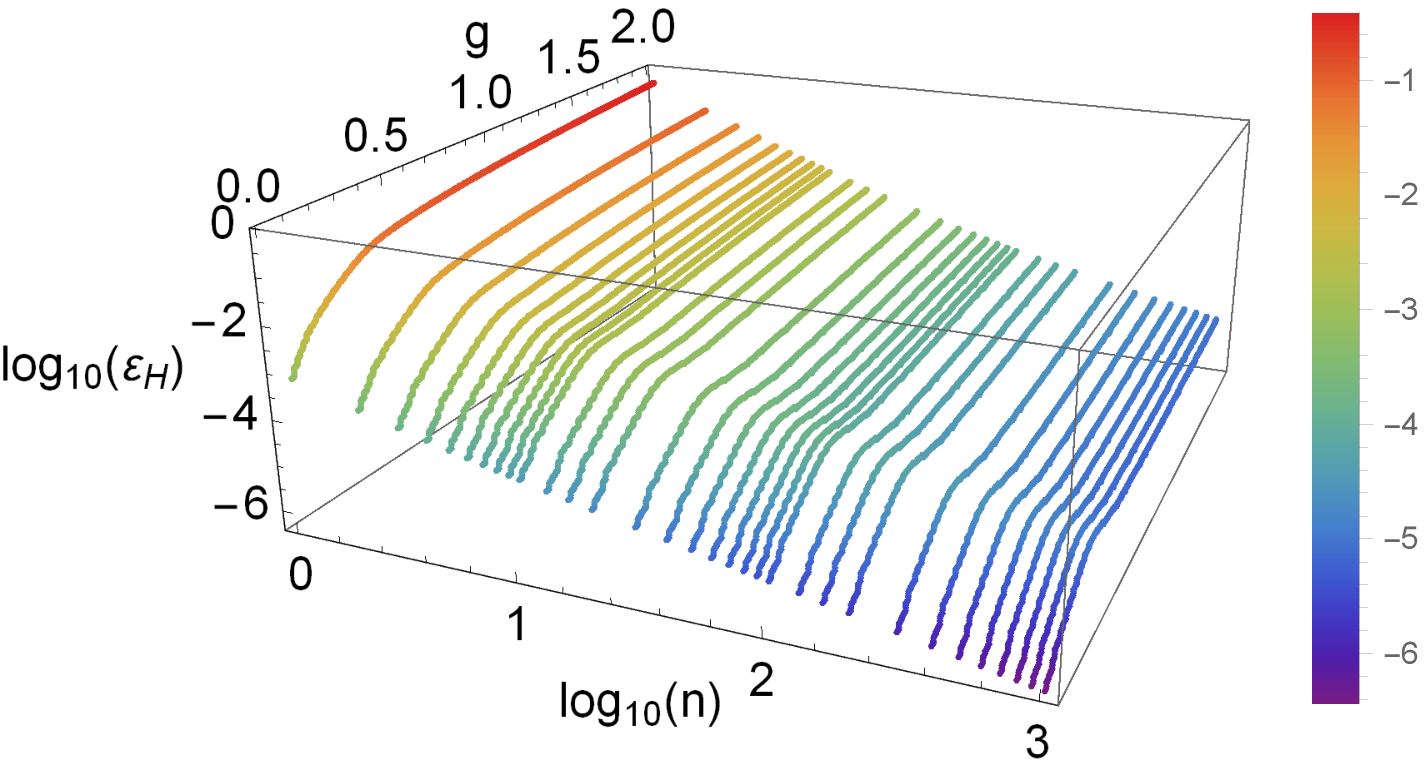}
      \subfigure[]{}
      \includegraphics[width=0.45\textwidth]{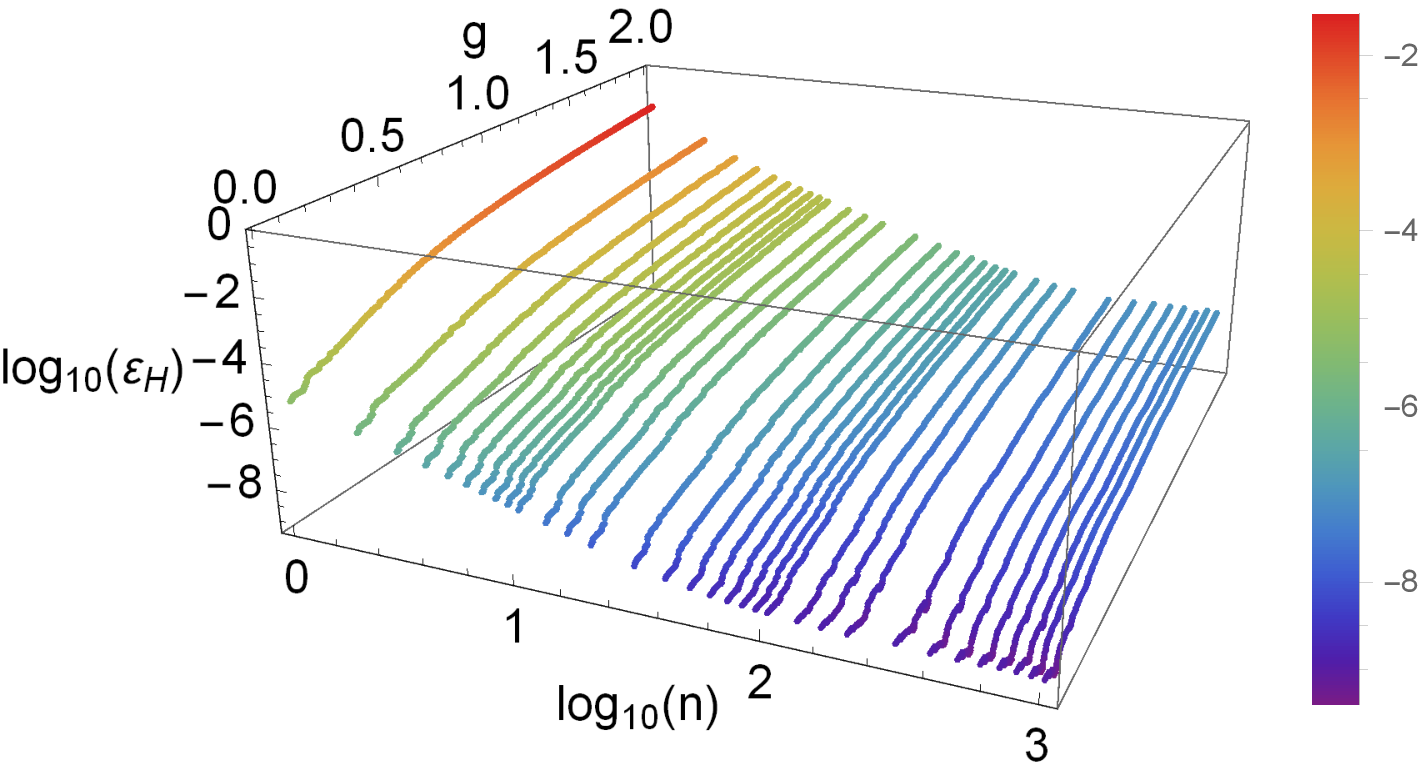}
     \caption{Numerical results of $log_{10}(\varepsilon_H)$ as a function of $g$  and $log_{10}(n)$. $g$ increases from $0$ to $2.0$ in steps of $g_s = 0.001$ and $t_s=0.1$, while $log_{10}(n)$  increases from $0$ to $3$. (a) Asymmetric Trotter decomposition. (b) Symmetric Trotter decomposition.
   }
      \label{fig_EB_3D}
\end{figure}

There are oscillations in the errors. So we define the    error bound of the energy   as
\begin{equation}
   \varepsilon_H(g) = \max \{|E_n(g')-E_0(g')|\},
\end{equation}
where the maximum is over
$$\ g- \frac{\Delta g}{2} \leqslant g' \leqslant g + \frac{\Delta g}{2}, $$
with $\Delta g$ representing a certain  window length. After some triers, we find $\Delta g= 0.04$ is about a small suitable value  to get ride of the oscillations. The error bounds $ \varepsilon_H(g)$ calculated from  $E_n(g)-E_0(g)$  in Fig.~\ref{fig_E}  are shown in Fig.~\ref{fig_smooth}. We have also calculated $ \varepsilon_H(g)$  for more values of $n$, which are shown  as functions of $g$ in  log-normal plots   in Fig.~\ref{fig_EB_n},   as functions of $n$ in  log-log  plots in Fig.~\ref{fig_EB_g},    and as functions of $n$ and $g$ in three-dimensional plots in   Fig.~\ref{fig_EB_3D}.

\begin{figure}[htb]
   \centering
      \subfigure[]{}
      \includegraphics[width=0.45\textwidth]{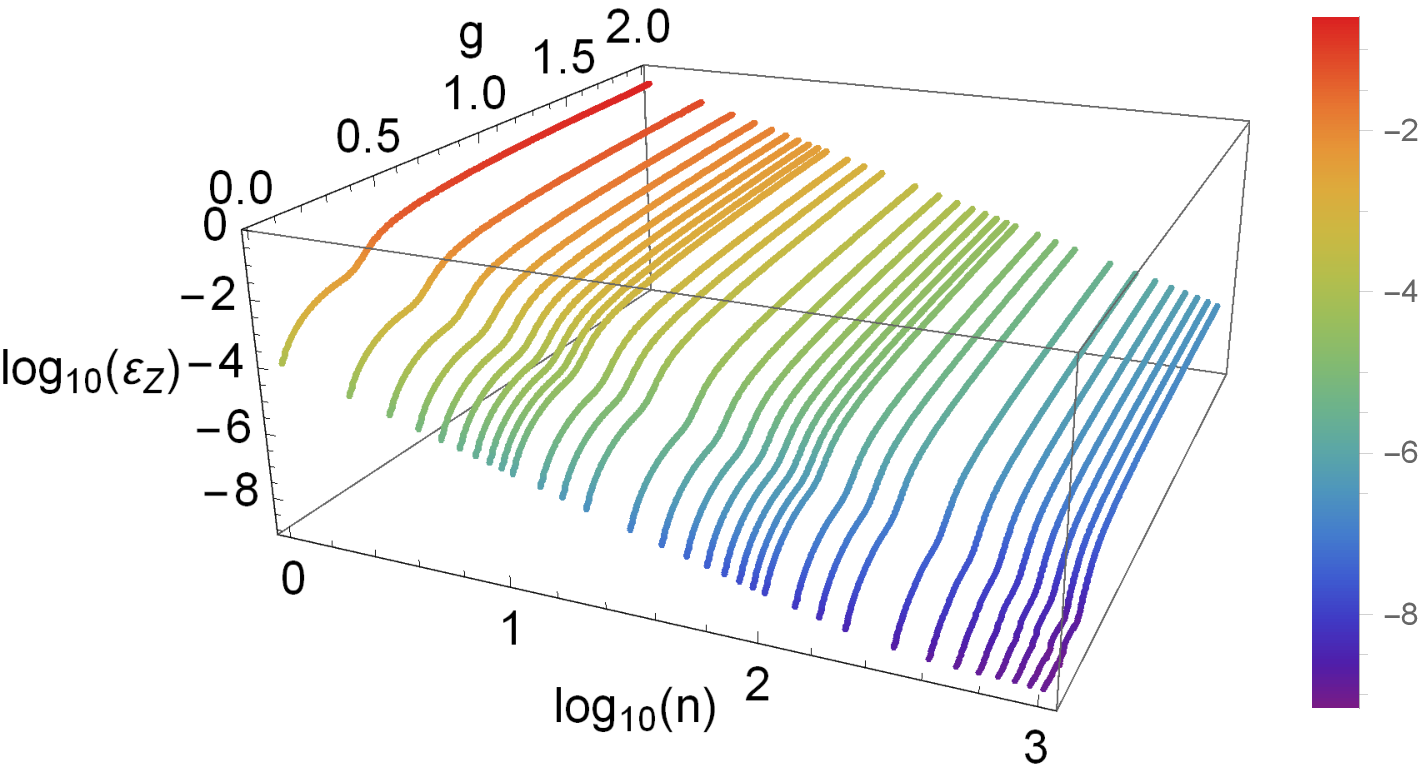}
      \subfigure[]{}
      \includegraphics[width=0.45\textwidth]{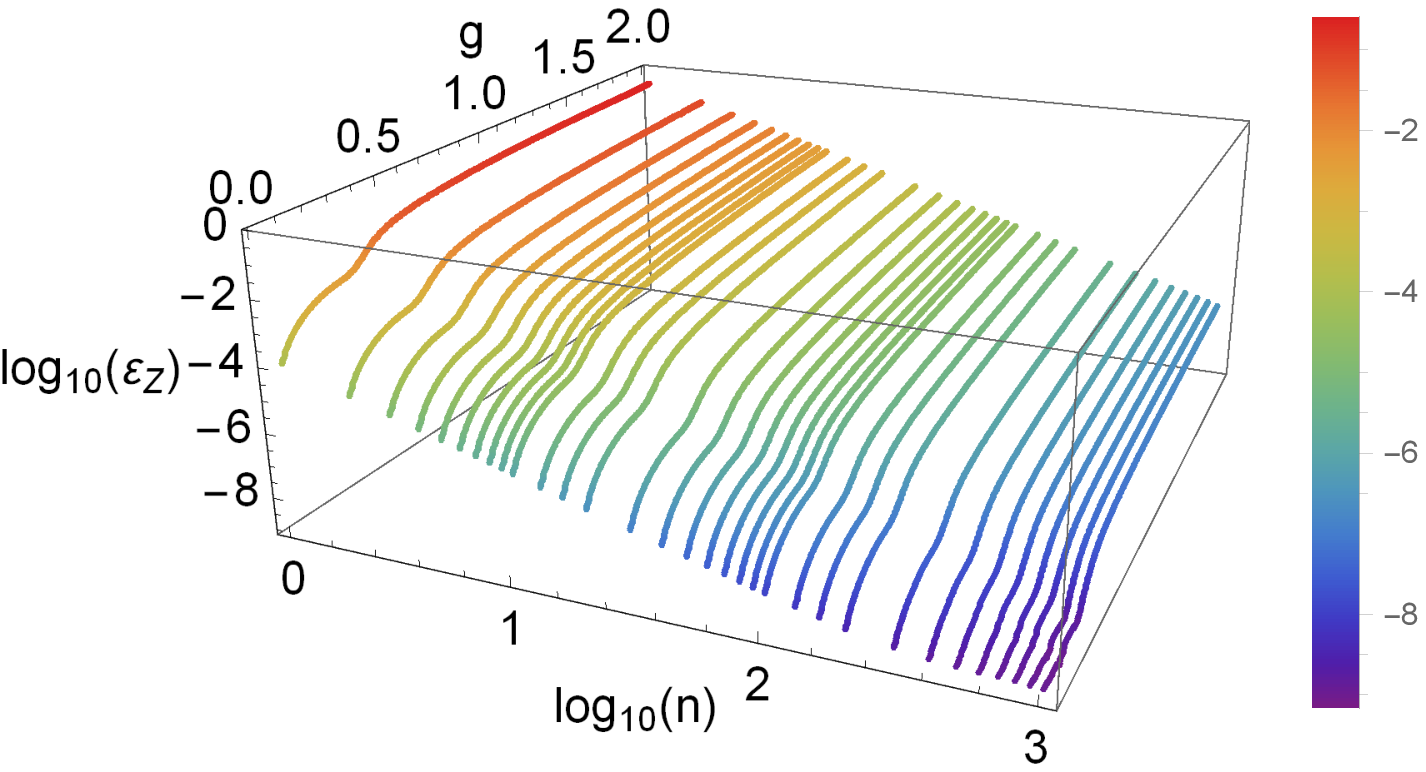}

      \subfigure[]{}
      \includegraphics[width=0.45\textwidth]{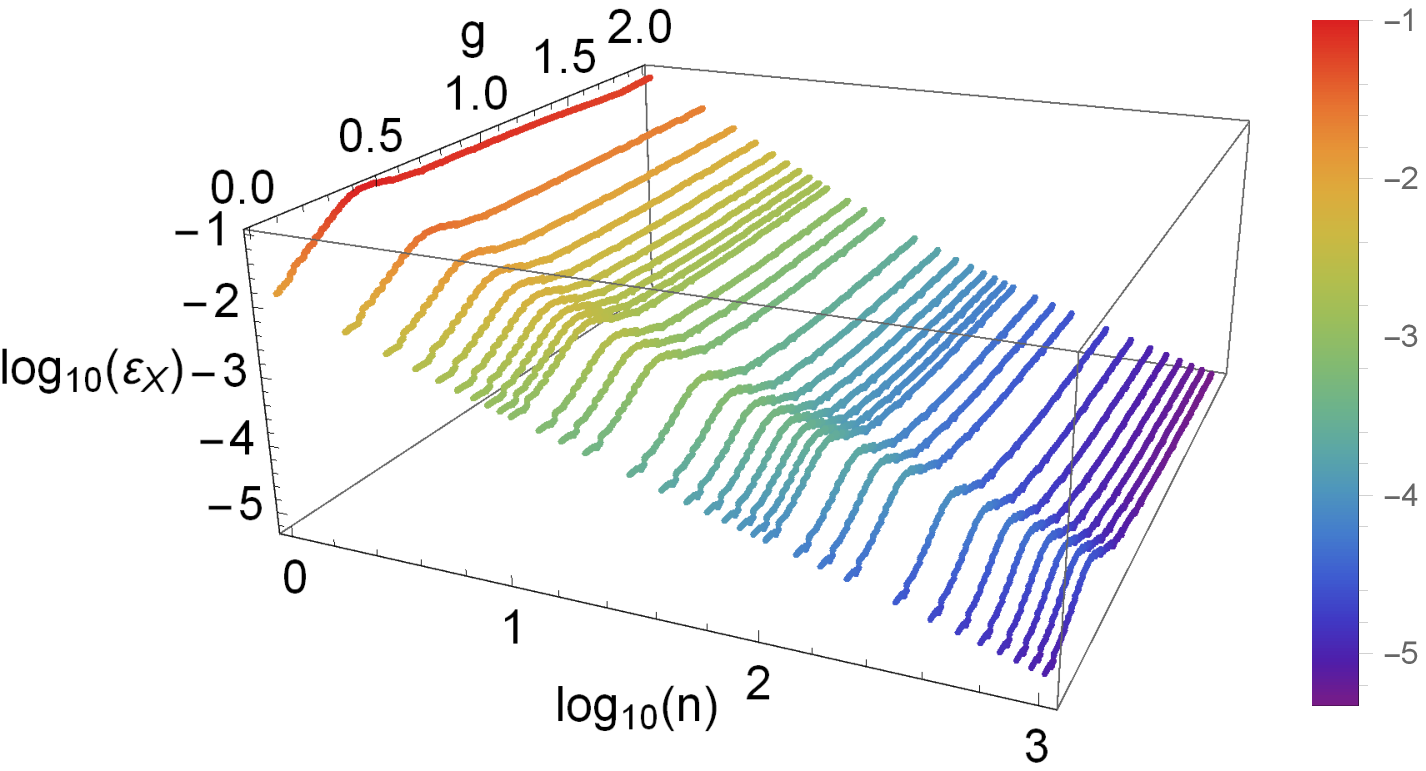}
      \subfigure[]{}
      \includegraphics[width=0.45\textwidth]{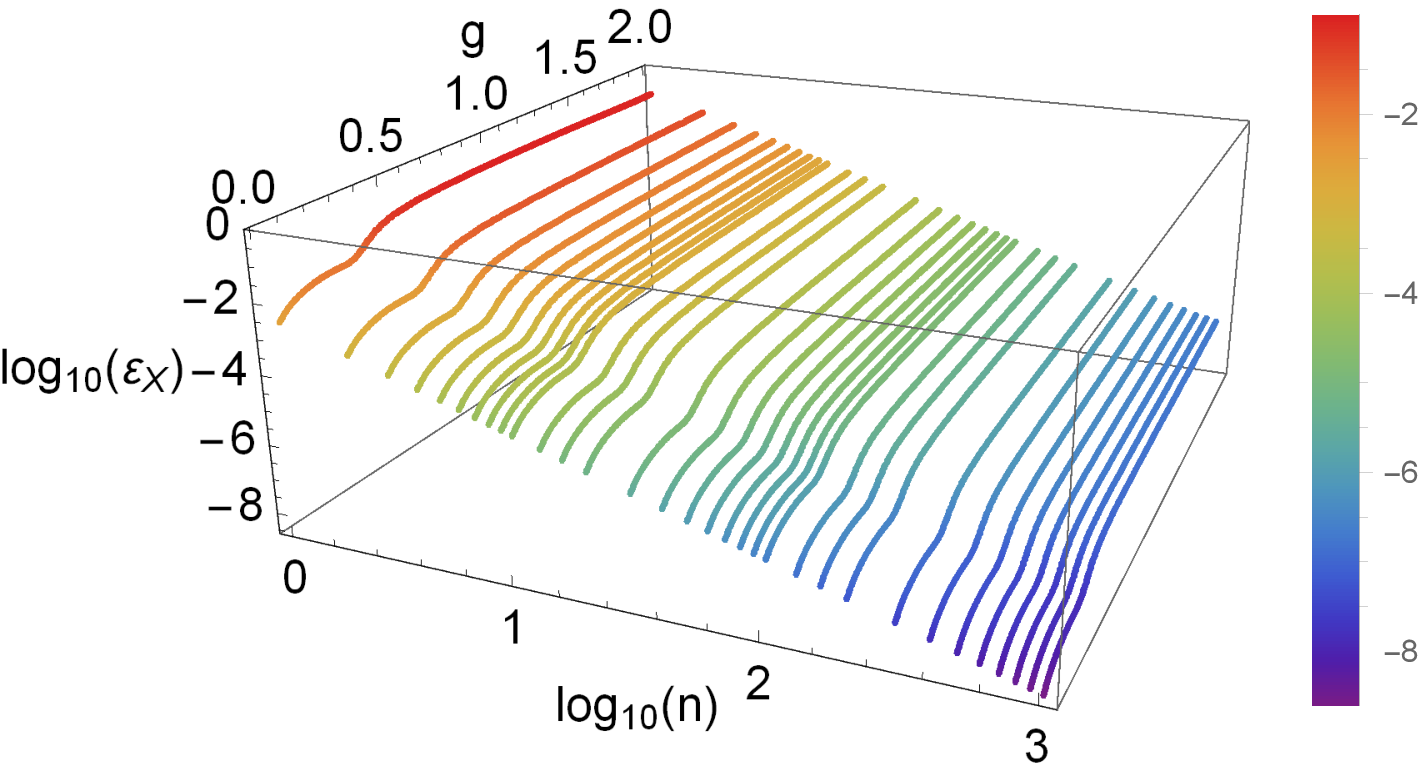}
   \caption{Numerical results of $log_{10}(\varepsilon_Z)$ and $log_{10}(\varepsilon_X)$ as functions of $g$,  which increases from $0$ to $2.0$ in step of $g_s = 0.001$ and $t_s=0.1$,  and $log_{10}(n)$,  which increases from $0$ to $3$. (a) $log_{10}(\varepsilon_Z)$ in the asymmetric Trotter decomposition. (b) $log_{10}(\varepsilon_Z)$ in the symmetric Trotter decomposition. (c) $log_{10}(\varepsilon_X)$ in the asymmetric Trotter decomposition.(d) $log_{10}(\varepsilon_X)$ in the symmetric Trotter decomposition.
   }
      \label{fig_ZXB_3D}
\end{figure}

Similarly,   the error bounds for  $Z$ and $X$ are defined  as
\begin{equation}
   \varepsilon_Z(g) = \max\{ | {\cal Z}_n(g') - {\cal Z}_0(g')|  \},
\end{equation}
and
\begin{equation}
   \varepsilon_X(g) = \max \{ | {\cal X}_n(g')-{\cal X}_0(g') | \},
\end{equation}
with the   window length for each case also found to be $0.04$.     $\varepsilon_Z$ and $\varepsilon_X$ as functions of $g$  and $n$ are shown in Fig.~\ref{fig_ZXB_3D}.

For  quantum $\mathbb{Z}_2$ LGT, there is a   QPT at $g_c \approx 0.38$. It can be seen from Fig.~\ref{fig_smooth}, Fig.~\ref{fig_EB_n} and Fig.~\ref{fig_EB_3D} that  for the asymmetric decomposition, the dependence of  $\varepsilon_H$ on $g$ exhibits a significant change  when $g$ is increased from $g<g_c$ to $g>g_c$, from a exponential   to  a  linear function.  For the symmetric  decomposition,  there is no such significant change, and $\varepsilon_H$ remains a   polynomial function of $g$.

It also can be seen from Fig.~\ref{fig_E} and Fig.~\ref{fig_ZXB_3D} that in  asymmetric and  symmetric decompositions,  $\varepsilon_Z$'s are  the same, for the reason given above about ${\cal Z}_n(g)-{\cal Z}_0(g)$,    but $\varepsilon_X$'s are quite different. For symmetric  decomposition, $\varepsilon_Z$ and $\varepsilon_X$ are close to each other at each value of $n$ and $g$, because  ${\cal Z}_n(g)-{\cal Z}_0(g)$ and ${\cal X}_n(g)-{\cal X}_0(g)$ are close in magnitude but opposite in sign, hence their cancellation   reduces $\varepsilon_H$.

In either  decomposition,  $\varepsilon_Z$ remains very small  when $g<g_c$,  and increases linearly with $g$ when $g>g_c$.  When $g<g_c$,  $\varepsilon_X$  increases exponentially with $g$ in   asymmetric     decomposition,   while  remains as small as  $\varepsilon_Z$ in symmetric decomposition.   When $g>g_c$,  in   asymmetric     decomposition,   $\varepsilon_X$  tends to be nearly unchanged, and thus  the    error bound  of $gX$   increases linearly with $g$,    while    in   symmetric     decomposition,  $\varepsilon_X$ remains close to  $\varepsilon_Z$.  These behaviors of errors and error bounds of  $Z$ and $X$ can explain that of $H$. Especially, within the parameter regime investigated here,   $\varepsilon_H$ in symmetric decomposition  is two orders of magnitude lower   than  that in asymmetric decomposition.

Fig.~\ref{fig_EB_g}, Fig.~\ref{fig_EB_3D} and Fig.~\ref{fig_ZXB_3D}  also indicate that in either decomposition,  $\varepsilon_H$,  $\varepsilon_Z$ and  $\varepsilon_X$ are almost  inversely proportional to $n$, in consistent with   the order-of-magnitude estimation of the errors.

\subsection{ Relative  errors}

We now turn to the relative   errors.

\begin{figure}[htb]
   \centering
   \subfigure[]{}
   \includegraphics[width=0.45\textwidth]{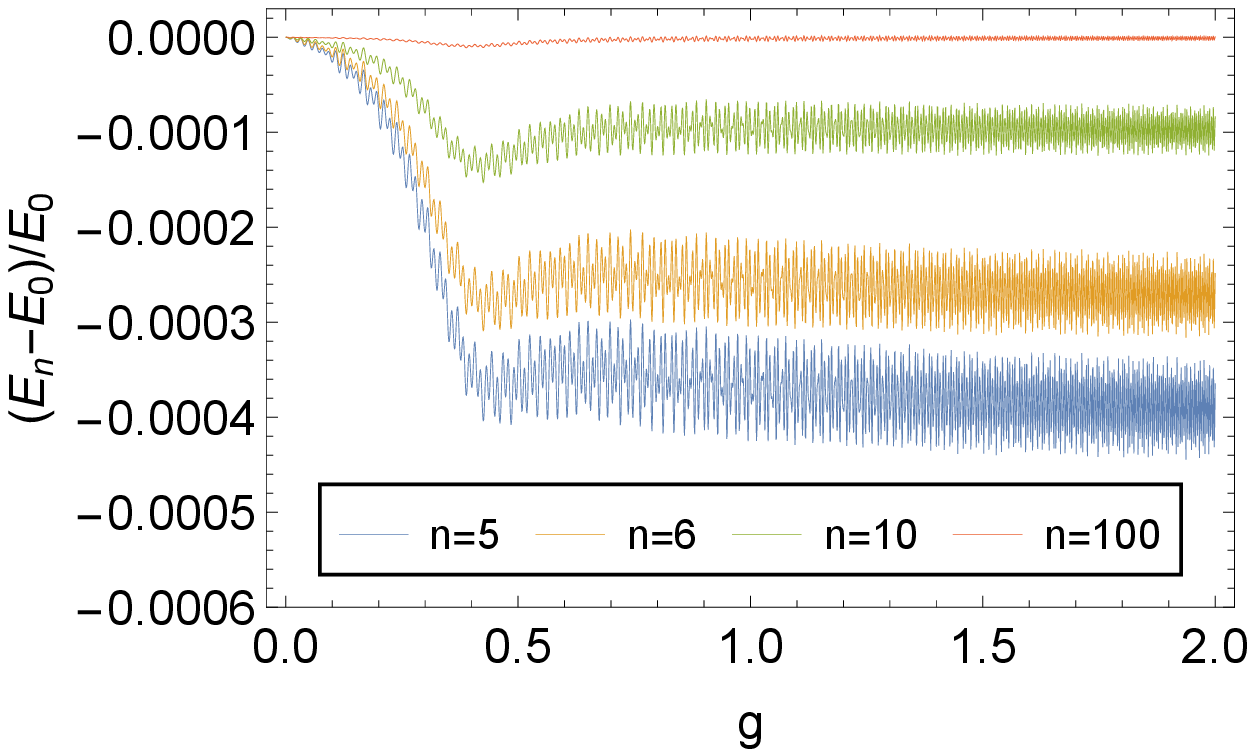}
   \subfigure[]{}
   \includegraphics[width=0.469\textwidth]{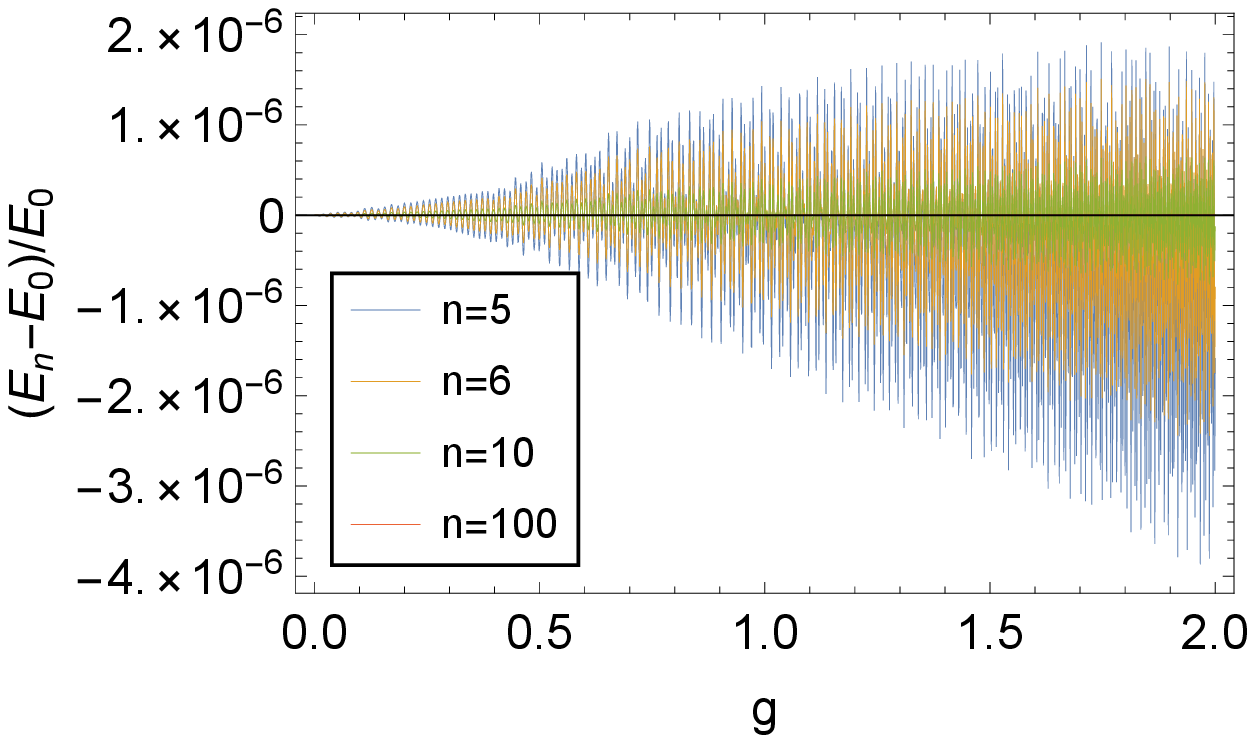}

\subfigure[]{}
   \includegraphics[width=0.45\textwidth]{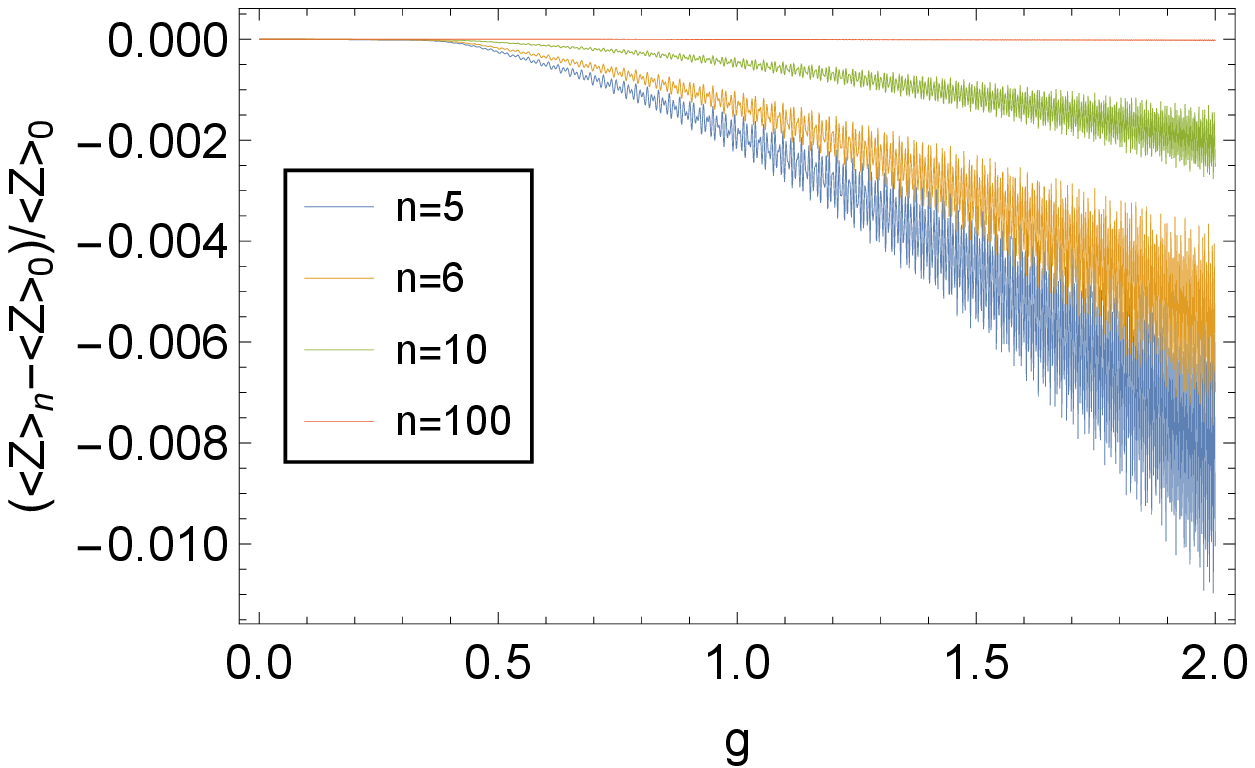}
   \subfigure[]{}
   \includegraphics[width=0.45\textwidth]{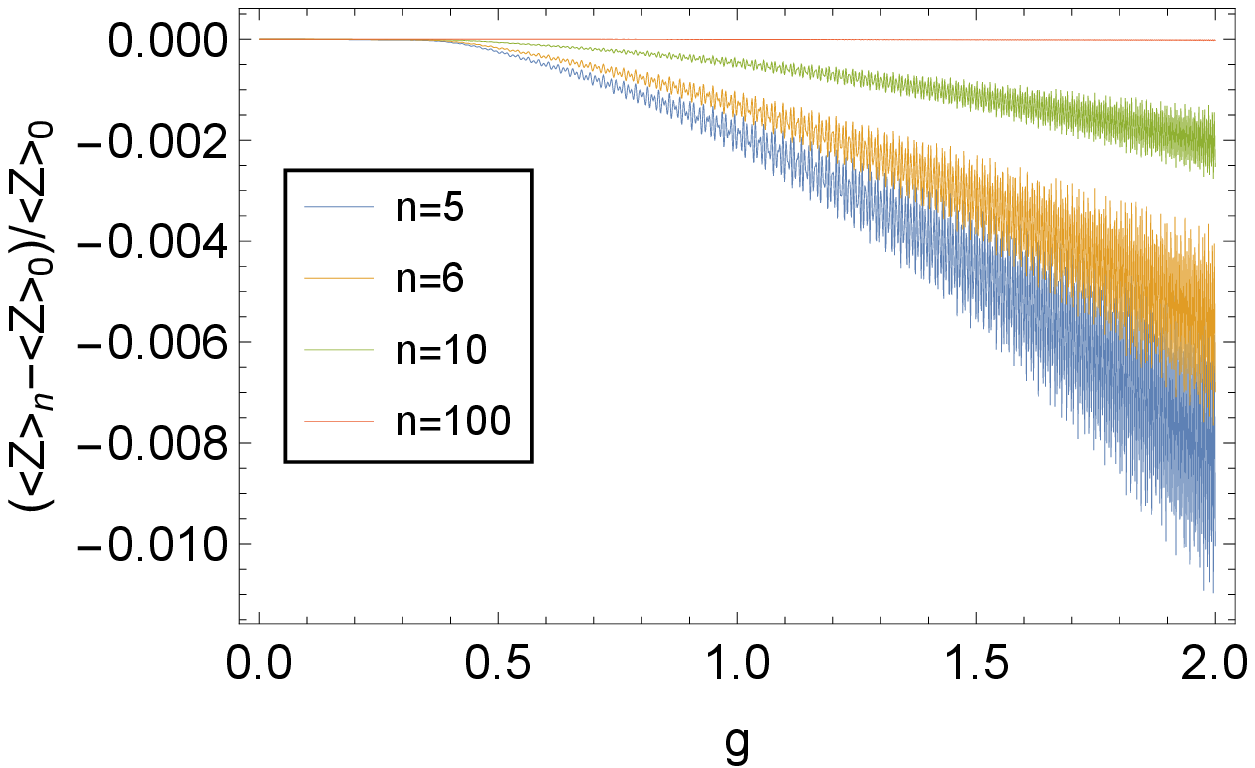}

   \subfigure[]{}
   \includegraphics[width=0.45\textwidth]{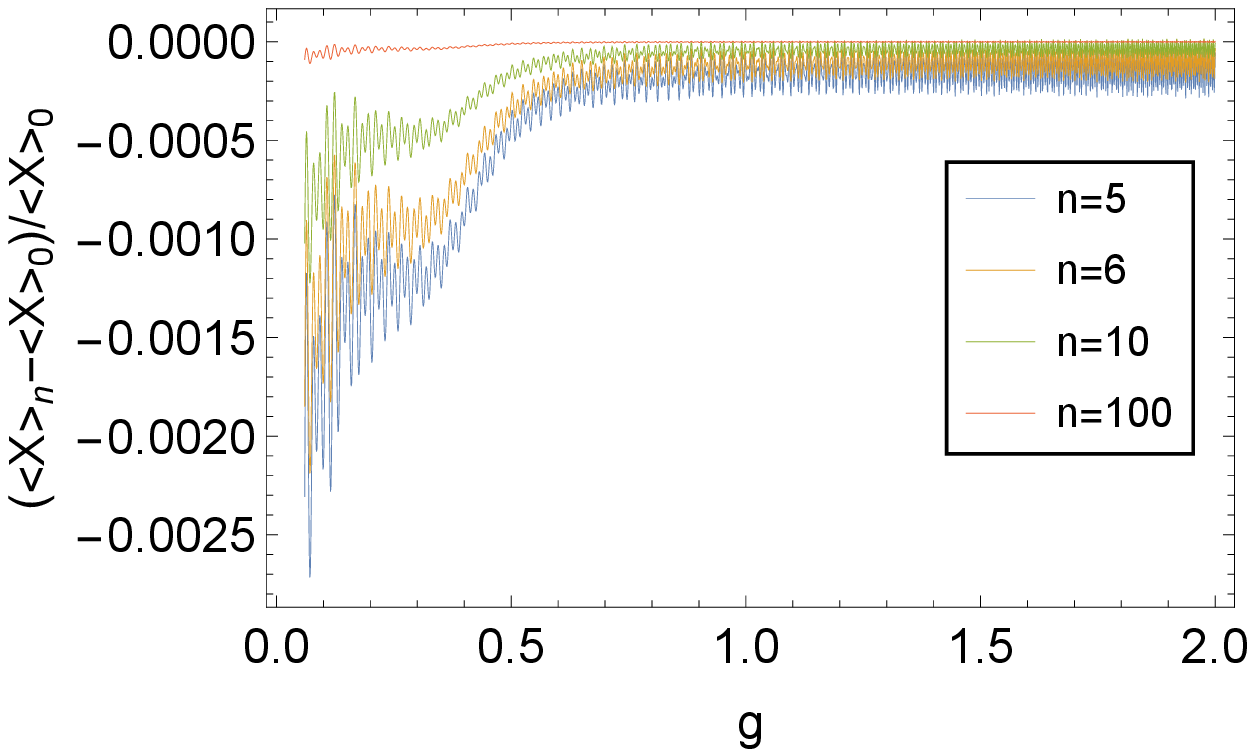}
   \subfigure[]{}
   \includegraphics[width=0.45\textwidth]{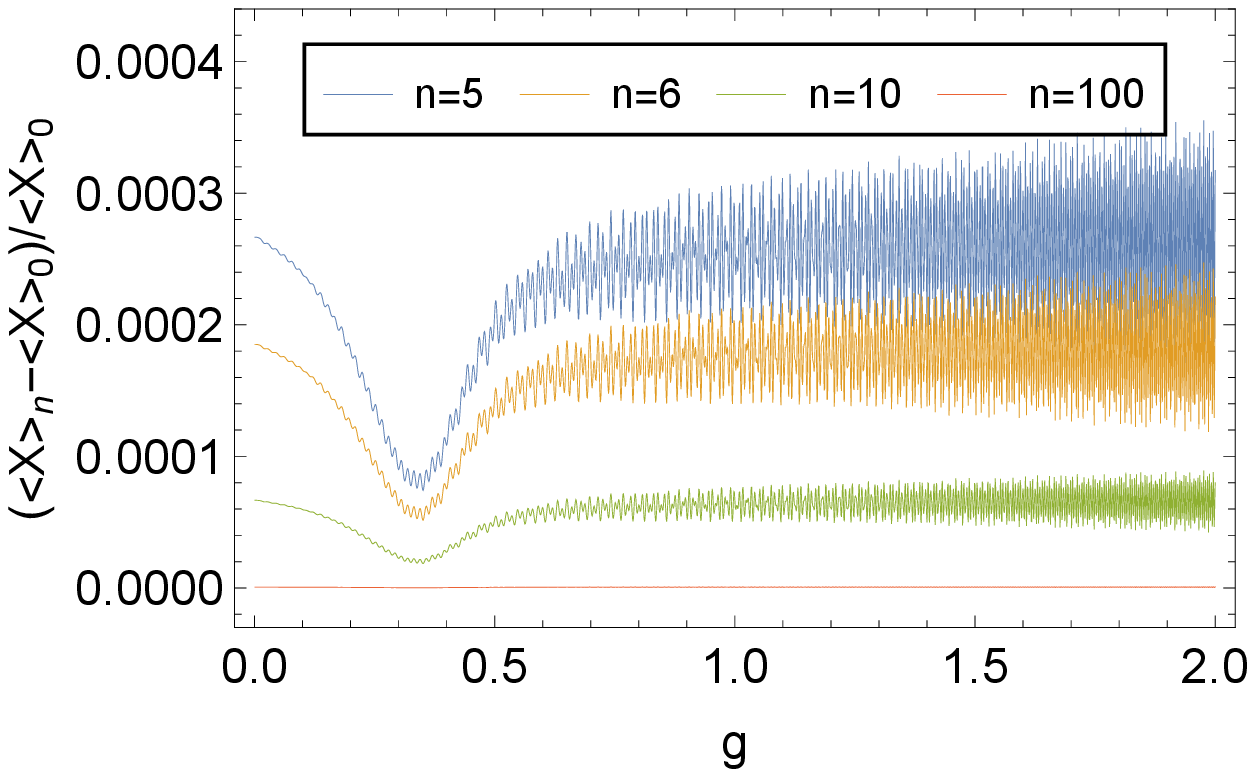}
   \caption{Numerical results of relative errors  as function of $g$, in steps of $g_s =0.001$ with  $n= 5, 6, 10, 100 $.  (a) $\left(E_n-E_0\right)/E_0$ in the asymmetric Trotter decomposition. (b) $\left(E_n-E_0\right)/E_0$ in the symmetric Trotter decomposition.  (c) $\left(\braket{Z}_n-\braket{Z}_0\right)/\braket{Z}_0$ in the  asymmetric Trotter decomposition. (d) $\left(\braket{Z}_n-\braket{Z}_0\right)/\braket{Z}_0$ in the symmetric Trotter decomposition. (e) $\left(\braket{X}_n-\braket{X}_0\right)/\braket{X}_0$ in the asymmetric Trotter decomposition. (f) $\left(\braket{X}_n-\braket{X}_0\right)/\braket{X}_0$ in the symmetric Trotter decomposition.}
   \label{fig_RE}
\end{figure}

$r(g)\equiv \left(E_n(g)-E_0(g)\right)/E_0(g)$ is shown in Fig.~\ref{fig_RE}, where it can be seen  that it is also two orders of magnitude lower  in the symmetric decomposition  than  in the  asymmetric decomposition. In the asymmetric decomposition, for $g>g_c$,  the relative error  $r(g)$ is nearly independent of $g$.

Fig.~\ref{fig_RE} also shows
$\left({\cal Z}_n(g)-{\cal Z}_0(g)\right)/{\cal Z}_0(g)$ and $\left({\cal X}_n(g)-{\cal X}_0(g)\right)/{\cal X}_0(g)$, with the latter about  one order of magnitude less. As ${\cal X}_0(g)$ is extremely small when $g$ is very small,  $\left({\cal X}_n(g)-{\cal X}_0(g)\right)/{\cal X}_0(g)$ is calculated for $g>0.06$.   $\left({\cal Z}_n(g)-{\cal Z}_0(g)\right)/{\cal Z}_0(g)$ in both decompositions are   nearly same. While $\left({\cal Z}_n(g)-{\cal Z}_0(g)\right)/{\cal Z}_0(g)$ is positive  $\left({\cal Z}_n(g)-{\cal Z}_0(g)\right)/{\cal Z}_0(g)$ is negative, as ${\cal Z}_0(g)$ is negative. Similarly, in asymmetric decomposition, $\left({\cal X}_n(g)-{\cal X}_0(g)\right)/{\cal X}_0(g)$ is also negative, hence $\left(E_n(g)-E_0(g)\right)/E_0(g)$ is the sum of two nagative numbers. In  symmetric decomposition,  $\left({\cal X}_n(g)-{\cal X}_0(g)\right)/{\cal X}_0(g)$ is   positive, hence $\left(E_n(g)-E_0(g))\right)/E_0(g)$ is a sum of one nagative number and one positive number. Consequently,  $\left(E_n(g)-E_0(g)\right)/E_0(g)$   is significantly smaller in the symmetric decomposition than in asymmetric decomposition.

\subsection{ Relative Error bounds  }

\begin{figure}[htb]
   \centering
   \subfigure[]{}
   \includegraphics[width=0.45\textwidth]{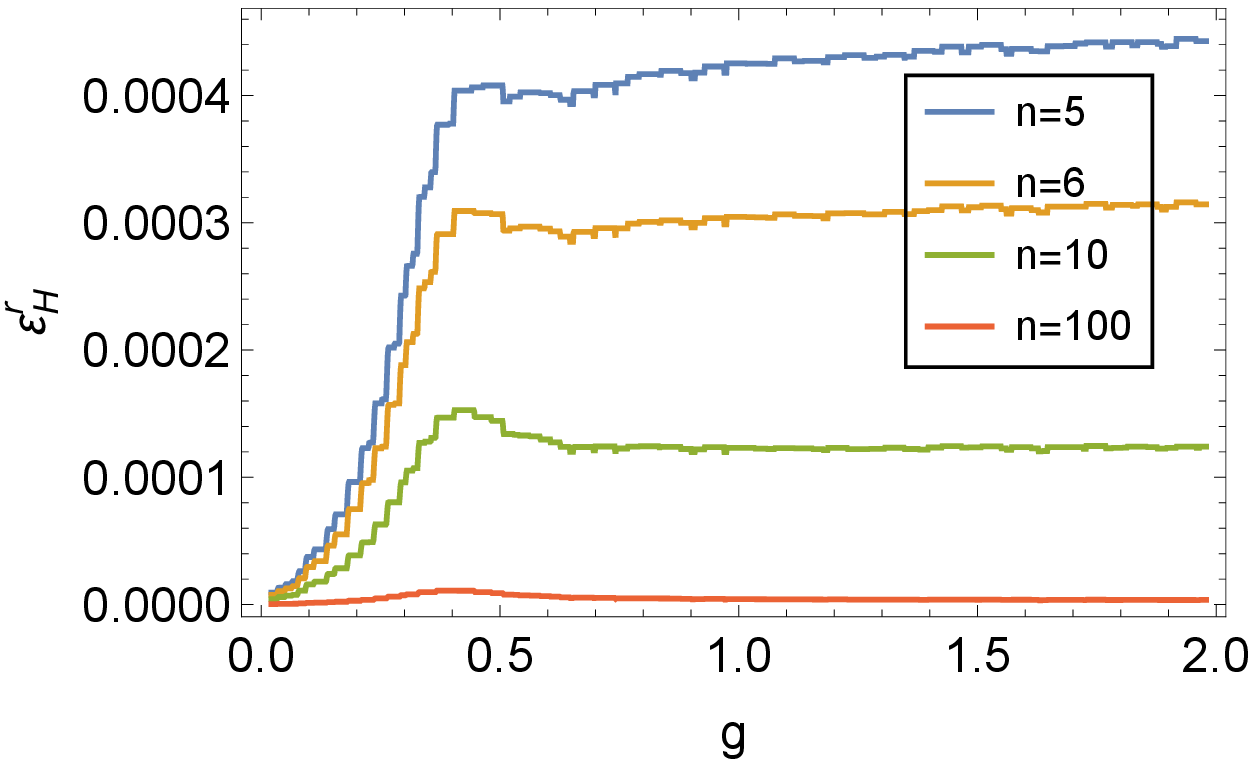}
   \subfigure[]{}
   \includegraphics[width=0.469\textwidth]{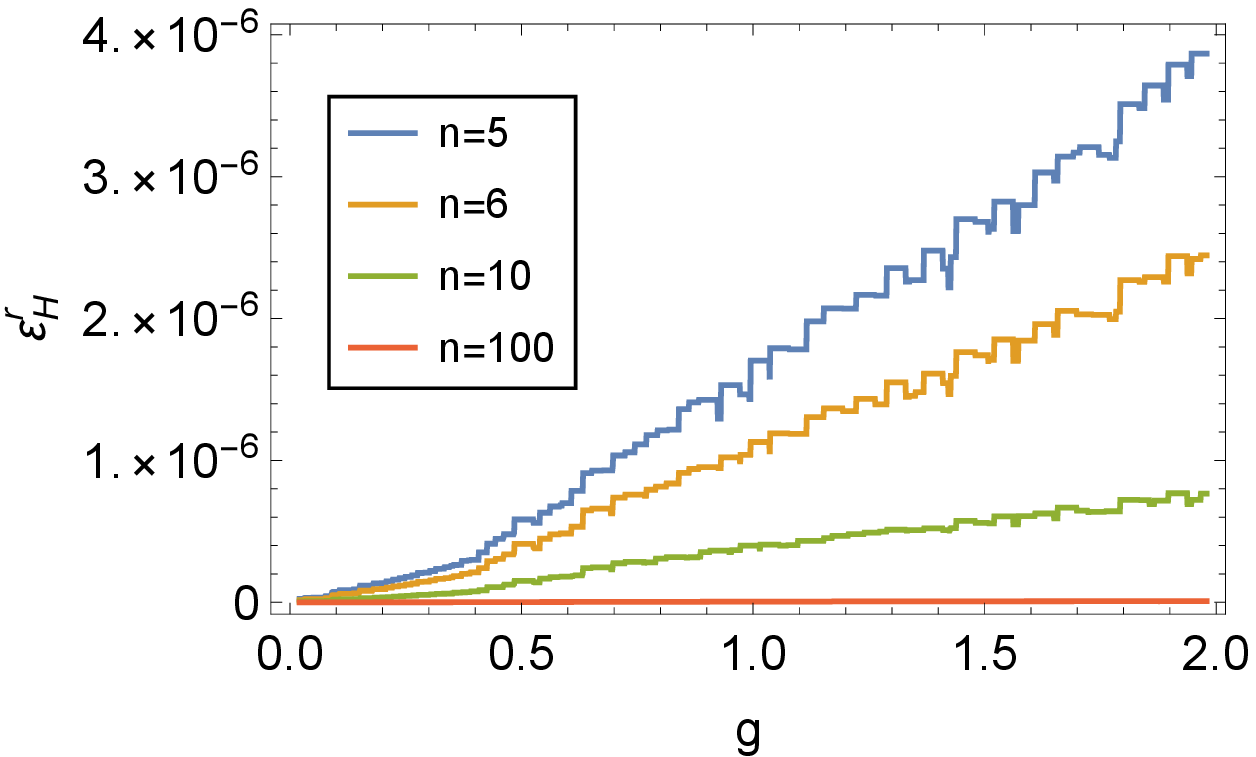}
    \caption{Relative error bound  $\varepsilon^r_H$  calculated from numerical results of $\left(E_n-E_0\right)/E_0$,  as a function of g, in step of $g_s =0.001$ with   $n=5, 6, 10, 100$. (a) Asymmetric Trotter decomposition. (b) Symmetric Trotter decomposition.   }\label{fig_Rsmooth}
\end{figure}

\begin{figure}[htb]
   \centering
      \subfigure[]{}
      \includegraphics[width=0.46\textwidth]{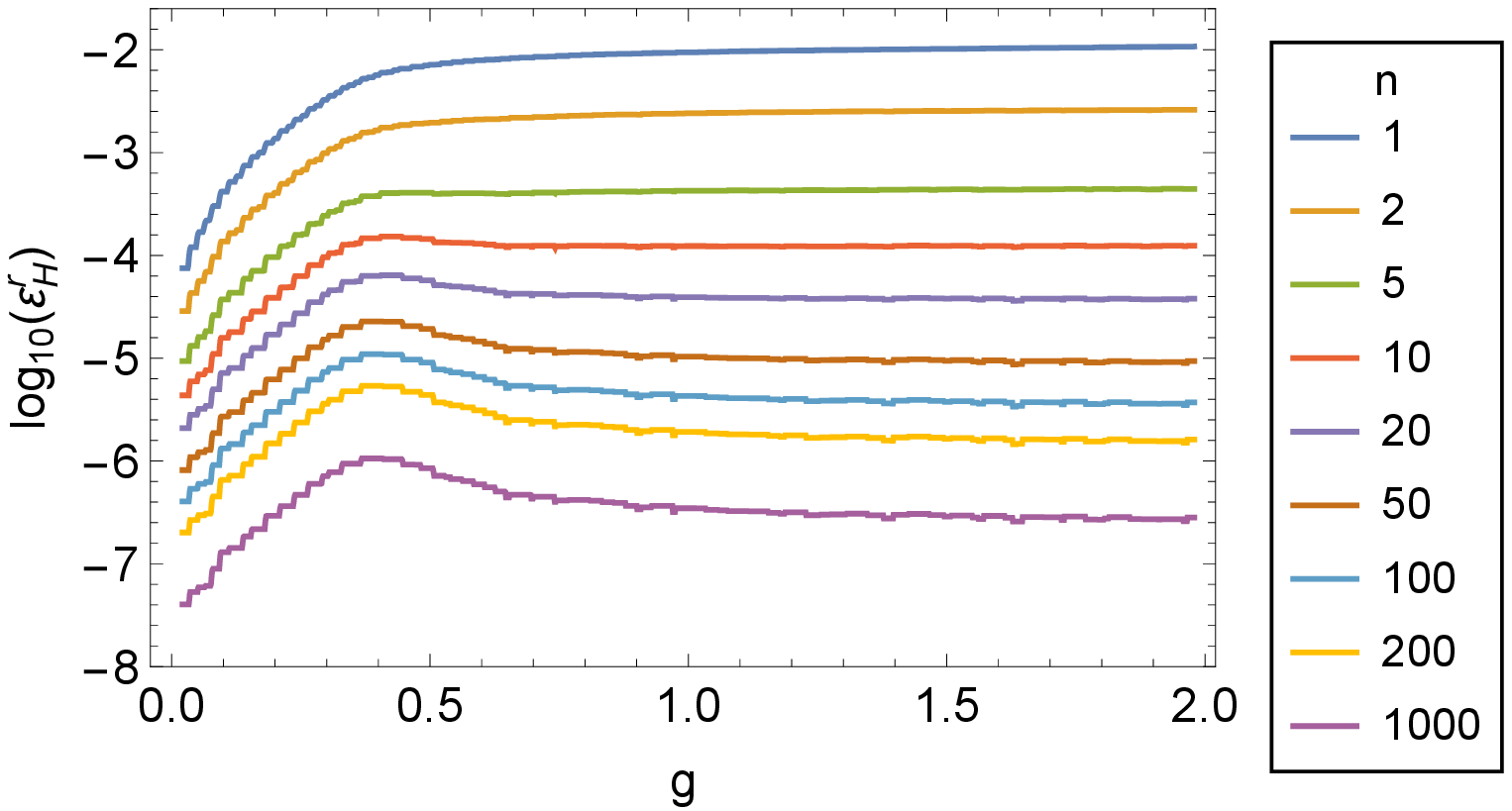}
      \subfigure[]{}
      \includegraphics[width=0.46\textwidth]{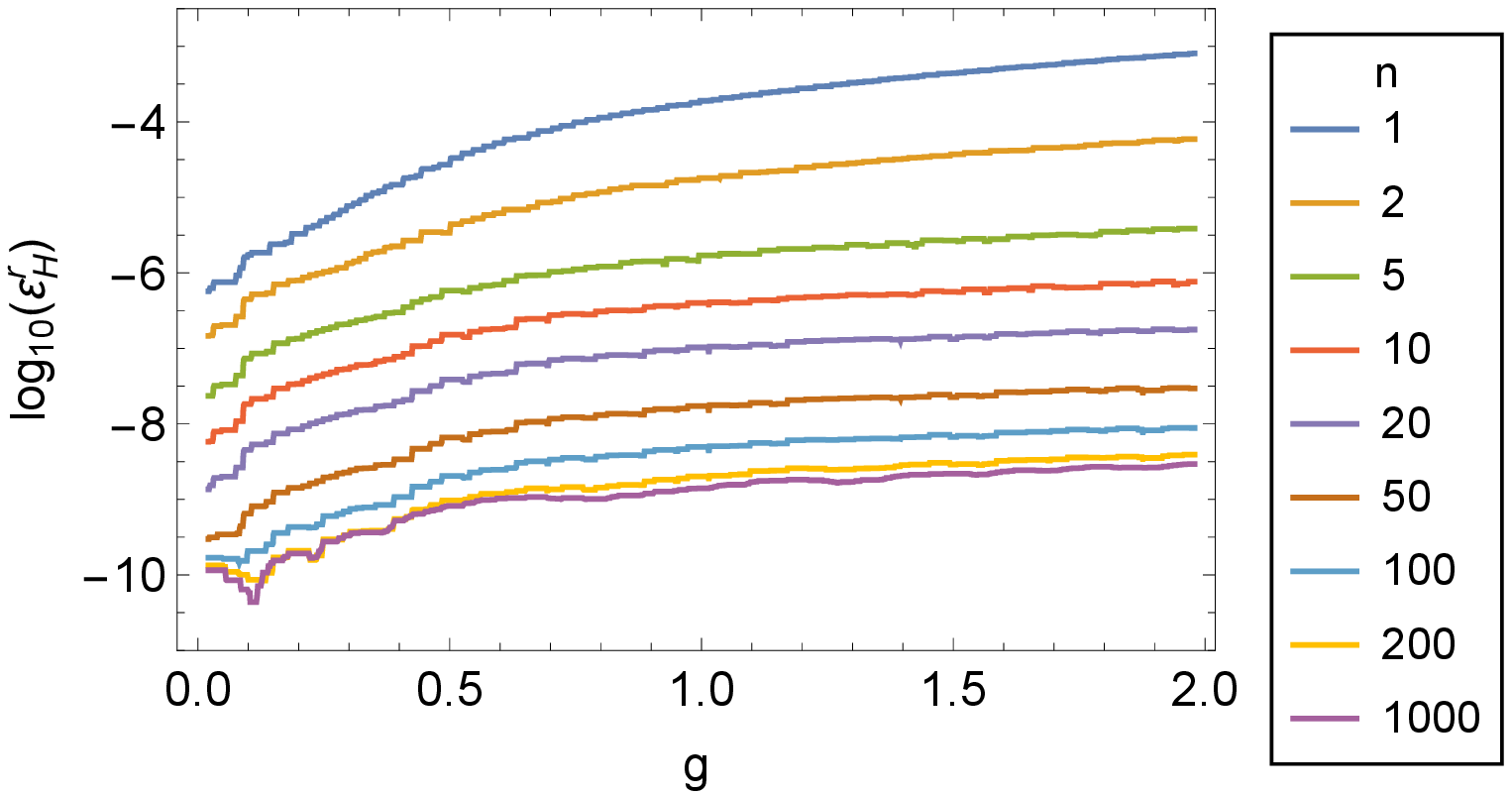}
      \caption{Numerical results of $log_{10}(\varepsilon^r_H)$ as functions of $g$, in steps of $g_s =0.001$ with various  values of  $n$. (a) Asymmetric Trotter decomposition. (b) Symmetric Trotter decomposition.}
      \label{fig_REB_n}
   \end{figure}

\begin{figure}[htb]
   \centering
      \subfigure[]{}
      \includegraphics[width=0.45\textwidth]{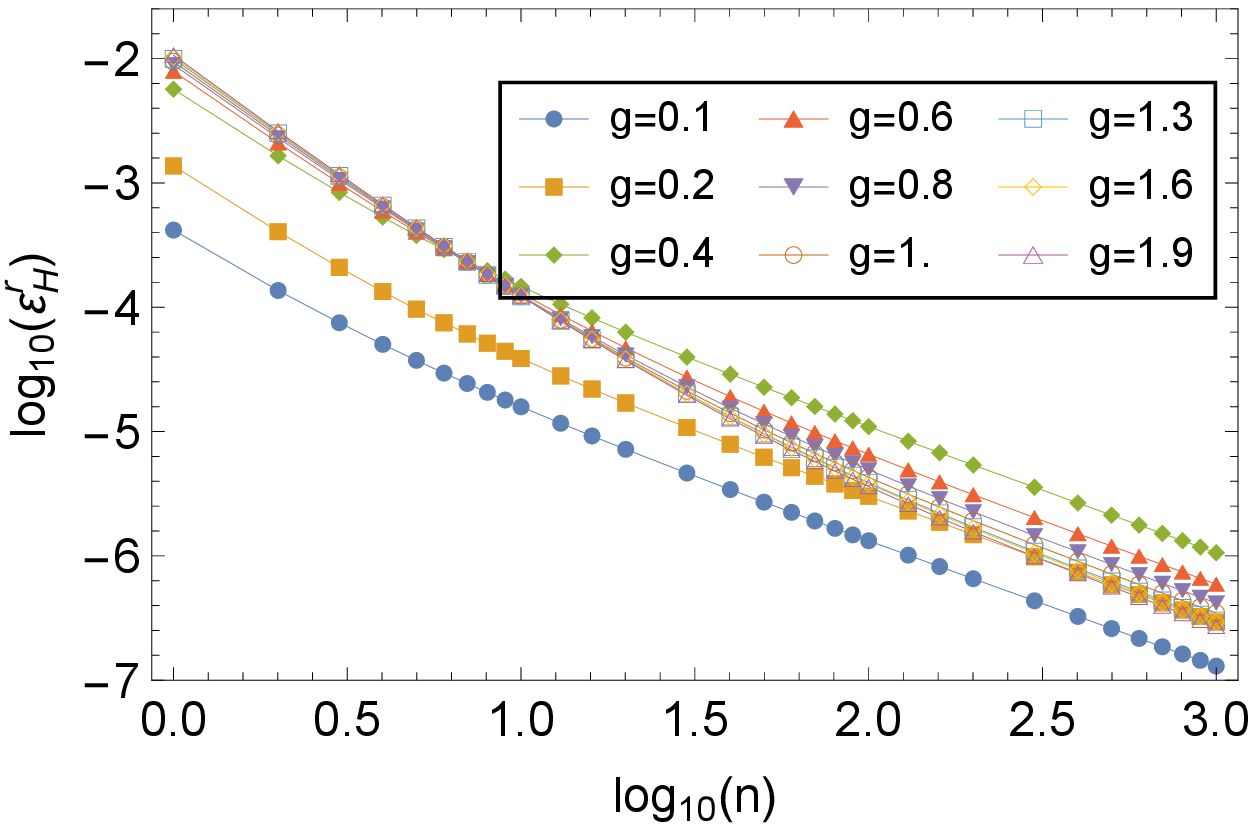}
      \subfigure[]{}
      \includegraphics[width=0.45\textwidth]{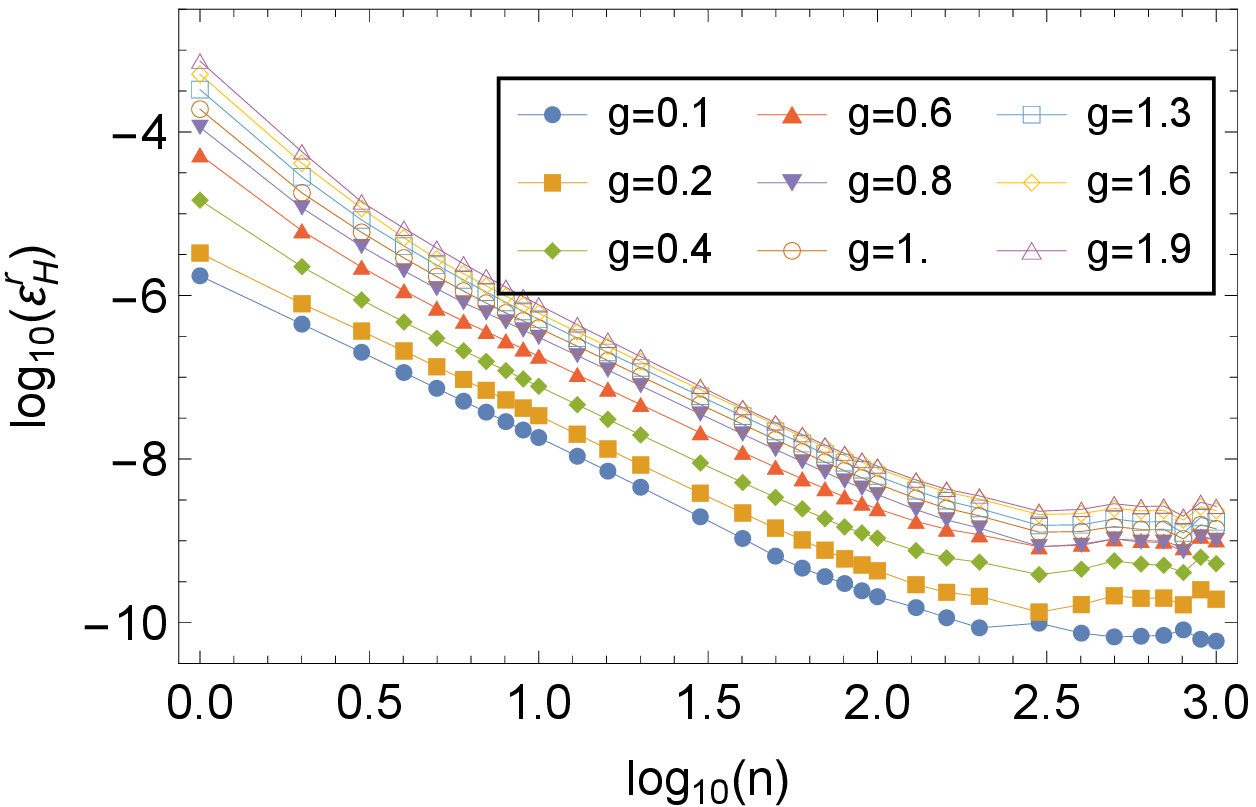}

      \caption{Numerical results of $log_{10}(\varepsilon^r_H)$ as functions of $n$ for various values of $g$. (a) Asymmetric Trotter decomposition. (b) Symmetric Trotter decomposition.
      }
      \label{fig_REB_g}
   \end{figure}

\begin{figure}[htb]
   \centering
      \subfigure[]{}
      \includegraphics[width=0.45\textwidth]{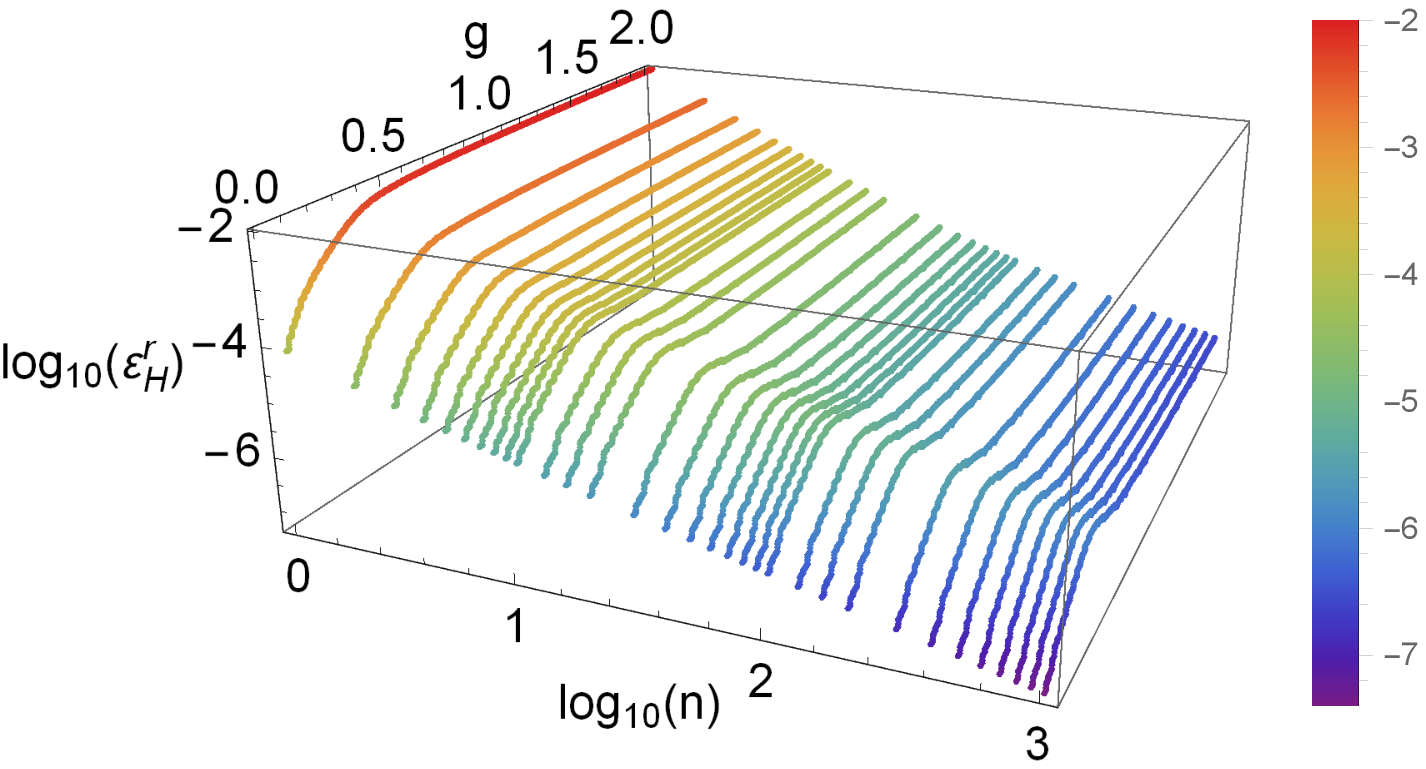}
      \subfigure[]{}
      \includegraphics[width=0.45\textwidth]{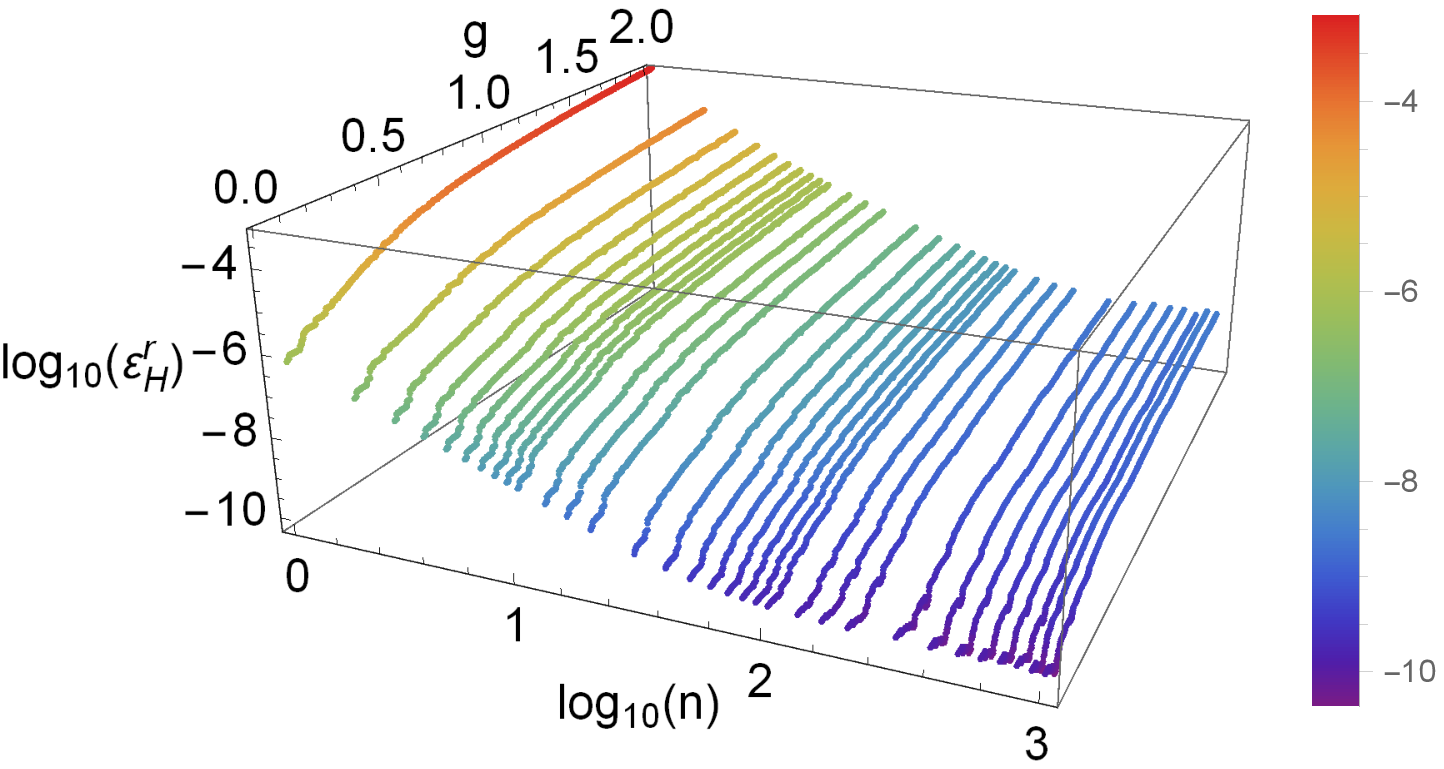}
     \caption{Numerical results of $log_{10}(\varepsilon^r_H)$ as a function of $g$  and $log_{10}(n)$. $g$ increases from $0$ to $2.0$ in steps of $g_s = 0.001$ and $t_s=0.1$, while $log_{10}(n)$  increases from $0$ to $3$. (a) Asymmetric Trotter decomposition. (b) Symmetric Trotter decomposition.
   }
      \label{fig_REB_3D}
\end{figure}

We define the   relative error bound of the energy   as
\begin{equation}
   \varepsilon^r_H(g) = \max \{\frac{|E_n(g')-E_0(g')|}{|E_0(g')|}  \},
\end{equation}
where the maximum is over
$$\ g- \frac{\Delta g}{2} \leqslant g' \leqslant g + \frac{\Delta g}{2}, $$
with $\Delta g$ representing a certain  window length, which is also $\Delta g= 0.04$. The error bounds $ \varepsilon^r_H(g)$ calculated from  $\left(E_n(g)-E_0(g)\right)/E_0(g)$  in Fig.~\ref{fig_RE}  are shown in Fig.~\ref{fig_Rsmooth}. We have also calculated $ \varepsilon^r_H(g)$  for more values of $n$, which are shown  as functions of $g$ in  log-normal plots   in Fig.~\ref{fig_REB_n},   as functions of $n$ in  log-log  plots in Fig.~\ref{fig_REB_g},    and as functions of $n$ and $g$ in three-dimensional plots in   Fig.~\ref{fig_REB_3D}.

Similarly,   the relative error bounds for  $Z$ and $X$ are defined  as
\begin{equation}
   \varepsilon^r_Z(g) = \max \{ \frac{ |{\cal Z}_n(g') - {\cal Z}_0(g')|}{ |{\cal Z}_0(g')| }  \},
\end{equation}
and
\begin{equation}
   \varepsilon^r_X(g) = \max\{ \frac{  |{\cal X}_n(g')-{\cal X}_0(g')|}{| {\cal X}_0(g')| } \},
\end{equation}
with the   window length for each case also being  $0.04$.     $\varepsilon^r_Z$ and $\varepsilon^r_X$ as functions of $g$  and $n$ are shown in Fig.~\ref{fig_ZXB_3D}.

\begin{figure}[htb]
   \centering
      \subfigure[]{}
      \includegraphics[width=0.45\textwidth]{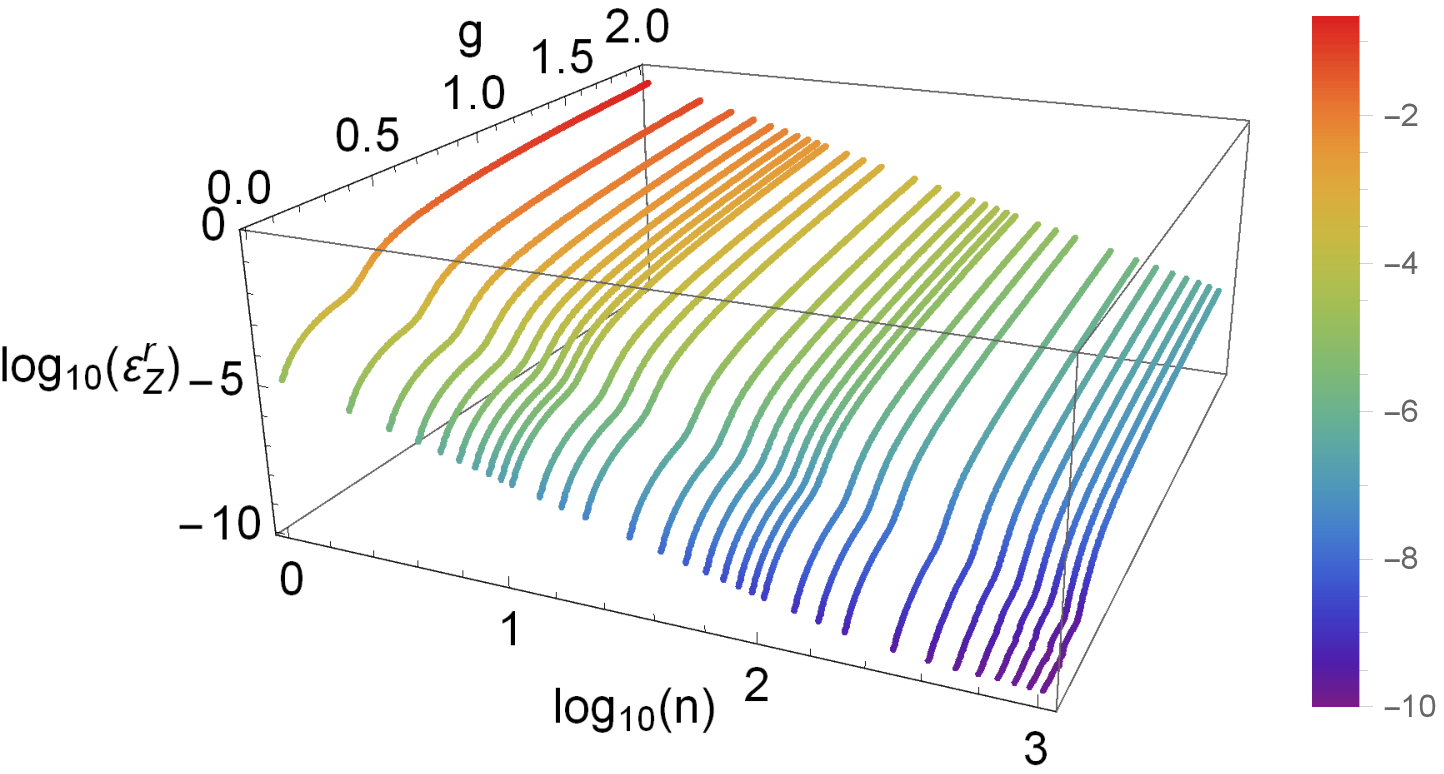}
      \subfigure[]{}
      \includegraphics[width=0.45\textwidth]{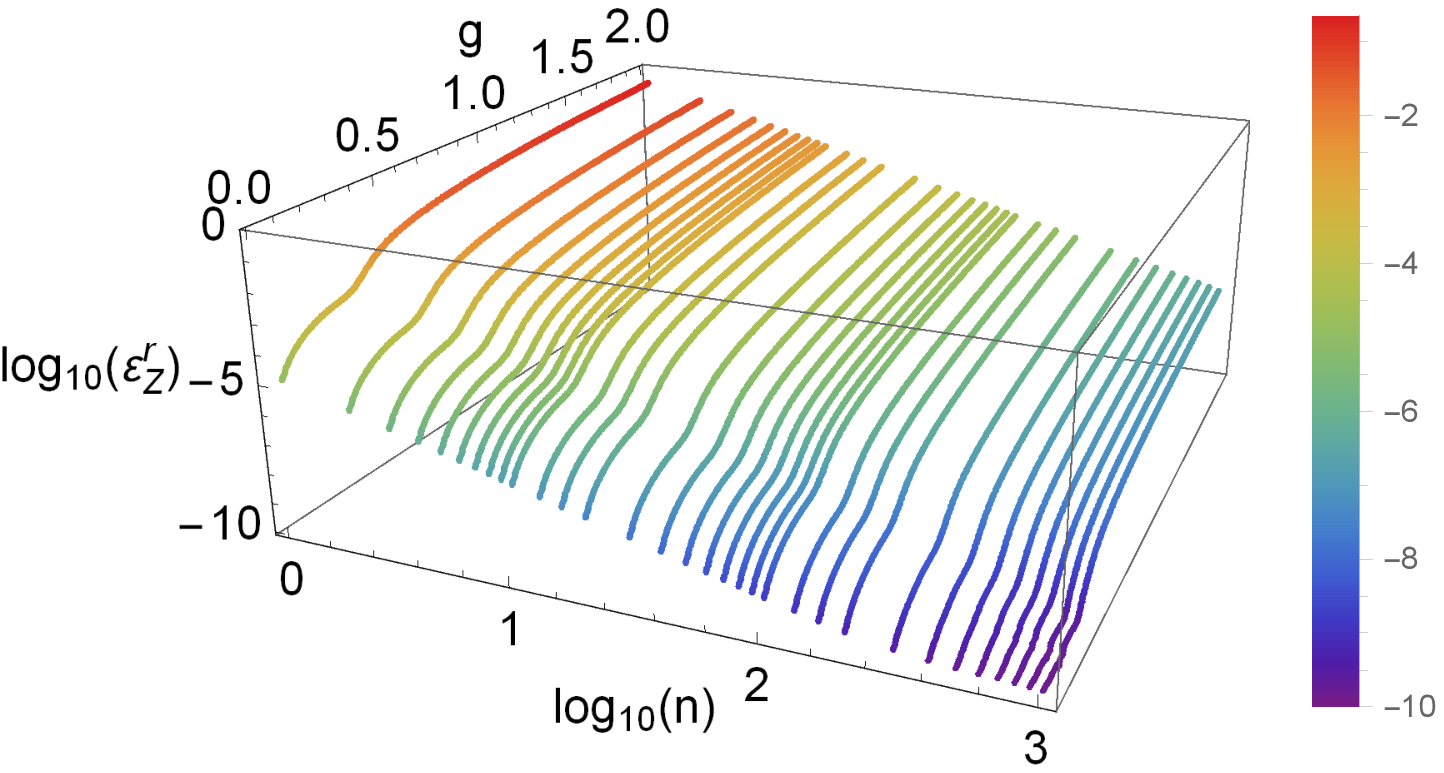}

      \subfigure[]{}
      \includegraphics[width=0.45\textwidth]{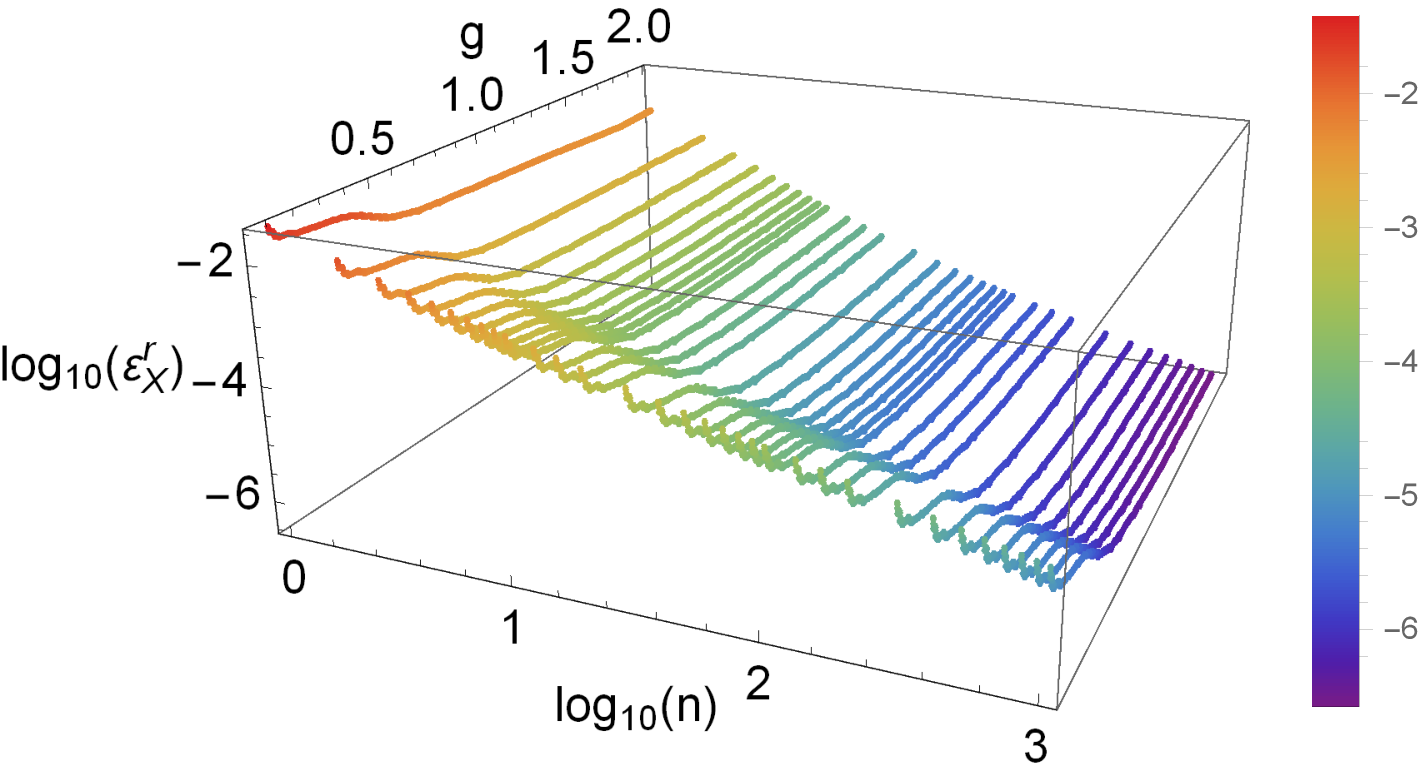}
      \subfigure[]{}
      \includegraphics[width=0.45\textwidth]{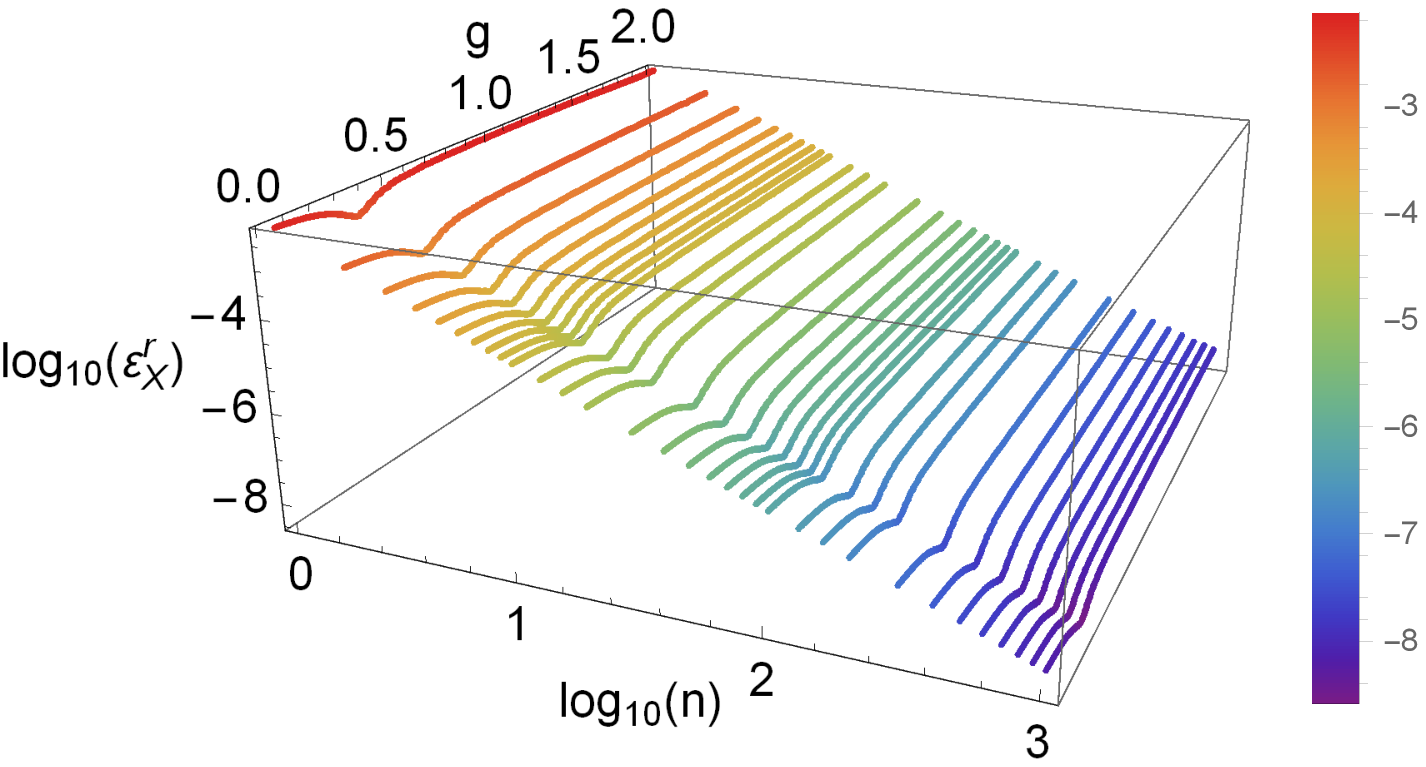}
   \caption{Numerical results of $log_{10}(\varepsilon^r_Z)$ and $log_{10}(\varepsilon^r_X)$ as functions of $g$,  which increases from $0$ to $2.0$ in step of $g_s = 0.001$ and $t_s=0.1$,  and $log_{10}(n)$,  which increases from $0$ to $3$. (a) $log_{10}(\varepsilon^r_Z)$ in the asymmetric Trotter decomposition. (b) $log_{10}(\varepsilon^r_Z)$ in the symmetric Trotter decomposition. (c) $log_{10}(\varepsilon^r_X)$ in the asymmetric Trotter decomposition.(d) $log_{10}(\varepsilon^r_X)$ in the symmetric Trotter decomposition.
   }
      \label{fig_RZXB_3D}
\end{figure}

It can also be seen from Fig.~\ref{fig_Rsmooth}, Fig.~\ref{fig_REB_n} and Fig.~\ref{fig_REB_3D} that  for the asymmetric decomposition, the dependence of  $\varepsilon^r_H$ on $g$ exhibits a significant change  when $g$ is increased from $g<g_c$ to $g>g_c$, namely, from an exponential increase to independence of $g$.
For the symmetric  decomposition,  there is no such a significant change.

It also can be seen from Fig.~\ref{fig_RE} and Fig.~\ref{fig_RZXB_3D} that in  asymmetric and  symmetric decompositions,  $\varepsilon^r_Z$'s are  the same,  but $\varepsilon^r_X$'s are quite different.

Fig.~\ref{fig_EB_g}, Fig.~\ref{fig_EB_3D} and Fig.~\ref{fig_ZXB_3D}  also indicate that in either decomposition,  $\varepsilon^r_H$,  $\varepsilon^r_Z$ and  $\varepsilon^r_X$ are almost  inversely proportional to $n$. This feature is shared by   order-of-magnitude estimation of the Trotter errors.

\section{Summary }

Trotter decomposition is the basis of digital quantum simulation. Here we have investigated the errors of symmetric and asymmetric Trotter decompositions and make comparisons between them, and with the order-of-magnitude estimation. Using a  GPU simulator, we have performed the numerical demonstration of digital adiabatic  quantum simulation of  quantum $\mathbb{Z}_2$ LGT, an approach called pseudoquantum simulation.

We have defined the errors in the energy, and  in expectation values of $Z$ and $X$, the two competing terms in the Hamiltonian. Each error is defined as the difference with the exact expectation value without the  Trotter error, which   is well approximated by using the Trotter decomposition with a very large number of steps. We have also defined the error bounds to get ride of small oscillations.

For symmetric and asymmetric  Trotter decomposition, and for various numbers of substeps, we calculate the errors as functions of the coupling  parameter $g$.

We observed clearly the  characteristic  differences  between    asymmetric and   symmetric Trotter decompositions. In the symmetric decomposition,  the errors in $\langle Z\rangle$ and  $\langle X\rangle$  are close in magnitude but opposite in sign, hence the  error and error bound in energy are about two orders of magnitude lower than in the  asymmetric decomposition.

In the asymmetric  and symmetric decomposition, errors and thus the error bounds of $\langle Z\rangle$ are the same, but those  of  $\langle X\rangle$ are different.

In the  asymmetric decomposition,  the error bound  $\varepsilon_H$    increases exponentially with $g$  for $g<g_c$, and   increases linearly  with $g$ for $g>g_c$.   In the symmetric decomposition,     $\varepsilon_H$  always   increases  with $g$  polynomially.

We have also investigated the relative errors and their bounds. Especially, we found that the relative error of energy is equal to the Trotter error defined in terms of the evolution operator.  The relative error bound of energy can be compared with the order-of-magnitude estimation for the Trotter errors, indicating that  the actual error is  much  lower than the  order-of-magnitude estimation,  especially when  $n$ is very small.

In each decomposition,   each actual  error is in  in inverse proportion to $n$. This relation is the same  as  in   the order-of-magnitude estimation.

These results provide useful information  for the experimental implementation of the  adiabatic quantum simulation of quantum $\mathbb{Z}_2$ LGT, and its pseudoquantum simulation.

\section{Acknowledge }

This work was supported by National Science Foundation of China (Grant No. 11574054).

\bibliography{qzt}

\begin{thebibliography}{23}%
\makeatletter
\providecommand \@ifxundefined [1]{%
 \@ifx{#1\undefined}
}%
\providecommand \@ifnum [1]{%
 \ifnum #1\expandafter \@firstoftwo
 \else \expandafter \@secondoftwo
 \fi
}%
\providecommand \@ifx [1]{%
 \ifx #1\expandafter \@firstoftwo
 \else \expandafter \@secondoftwo
 \fi
}%
\providecommand \natexlab [1]{#1}%
\providecommand \enquote  [1]{``#1''}%
\providecommand \bibnamefont  [1]{#1}%
\providecommand \bibfnamefont [1]{#1}%
\providecommand \citenamefont [1]{#1}%
\providecommand \href@noop [0]{\@secondoftwo}%
\providecommand \href [0]{\begingroup \@sanitize@url \@href}%
\providecommand \@href[1]{\@@startlink{#1}\@@href}%
\providecommand \@@href[1]{\endgroup#1\@@endlink}%
\providecommand \@sanitize@url [0]{\catcode `\\12\catcode `\$12\catcode
  `\&12\catcode `\#12\catcode `\^12\catcode `\_12\catcode `\%12\relax}%
\providecommand \@@startlink[1]{}%
\providecommand \@@endlink[0]{}%
\providecommand \url  [0]{\begingroup\@sanitize@url \@url }%
\providecommand \@url [1]{\endgroup\@href {#1}{\urlprefix }}%
\providecommand \urlprefix  [0]{URL }%
\providecommand \Eprint [0]{\href }%
\providecommand \doibase [0]{http://dx.doi.org/}%
\providecommand \selectlanguage [0]{\@gobble}%
\providecommand \bibinfo  [0]{\@secondoftwo}%
\providecommand \bibfield  [0]{\@secondoftwo}%
\providecommand \translation [1]{[#1]}%
\providecommand \BibitemOpen [0]{}%
\providecommand \bibitemStop [0]{}%
\providecommand \bibitemNoStop [0]{.\EOS\space}%
\providecommand \EOS [0]{\spacefactor3000\relax}%
\providecommand \BibitemShut  [1]{\csname bibitem#1\endcsname}%
\let\auto@bib@innerbib\@empty
\bibitem [{\citenamefont {Wegner}(1971)}]{Wegner1971}%
  \BibitemOpen
  \bibfield  {author} {\bibinfo {author} {\bibfnamefont {F.~J.}\ \bibnamefont
  {Wegner}},\ }\bibfield  {title} {\enquote {\bibinfo {title} {Duality in
  generalized ising models and phase transitions without local order
  parameters},}\ }\href {\doibase 10.1063/1.1665530} {\bibfield  {journal}
  {\bibinfo  {journal} {J. Math. Phys.}\ }\textbf {\bibinfo {volume} {12}},\
  \bibinfo {pages} {2259} (\bibinfo {year} {1971})}\BibitemShut {NoStop}%
\bibitem [{\citenamefont {Kogut}(1979)}]{Kogut1979}%
  \BibitemOpen
  \bibfield  {author} {\bibinfo {author} {\bibfnamefont {J.~B.}\ \bibnamefont
  {Kogut}},\ }\bibfield  {title} {\enquote {\bibinfo {title} {An introduction
  to lattice gauge theory and spin systems},}\ }\href {\doibase
  10.1103/RevModPhys.51.659} {\bibfield  {journal} {\bibinfo  {journal} {Rev.
  Mod. Phys.}\ }\textbf {\bibinfo {volume} {51}},\ \bibinfo {pages} {659}
  (\bibinfo {year} {1979})}\BibitemShut {NoStop}%
\bibitem [{\citenamefont {Sachdev}(2018)}]{sachdev}%
  \BibitemOpen
  \bibfield  {author} {\bibinfo {author} {\bibfnamefont {S.}~\bibnamefont
  {Sachdev}},\ }\bibfield  {title} {\enquote {\bibinfo {title} {Topological
  order, emergent gauge fields, and fermi surface reconstruction},}\ }\href
  {\doibase 10.1088/1361-6633/aae110} {\bibfield  {journal} {\bibinfo
  {journal} {Rep. Prog. Phys.}\ }\textbf {\bibinfo {volume} {82}},\ \bibinfo
  {pages} {014001} (\bibinfo {year} {2018})}\BibitemShut {NoStop}%
\bibitem [{\citenamefont {Kogut}\ and\ \citenamefont
  {Susskind}(1975)}]{Kogut_1975}%
  \BibitemOpen
  \bibfield  {author} {\bibinfo {author} {\bibfnamefont {J.}~\bibnamefont
  {Kogut}}\ and\ \bibinfo {author} {\bibfnamefont {L.}~\bibnamefont
  {Susskind}},\ }\bibfield  {title} {\enquote {\bibinfo {title} {{Hamiltonian
  formulation of Wilson's lattice gauge theories}},}\ }\href {\doibase
  10.1103/PhysRevD.11.395} {\bibfield  {journal} {\bibinfo  {journal} {Phys.
  Rev. D}\ }\textbf {\bibinfo {volume} {11}},\ \bibinfo {pages} {395--408}
  (\bibinfo {year} {1975})}\BibitemShut {NoStop}%
\bibitem [{\citenamefont {Wilson}(1974)}]{Wilson_1974}%
  \BibitemOpen
  \bibfield  {author} {\bibinfo {author} {\bibfnamefont {K.~G.}\ \bibnamefont
  {Wilson}},\ }\bibfield  {title} {\enquote {\bibinfo {title} {Confinement of
  quarks},}\ }\href {\doibase 10.1103/PhysRevD.10.2445} {\bibfield  {journal}
  {\bibinfo  {journal} {Phys. Rev. D}\ }\textbf {\bibinfo {volume} {10}},\
  \bibinfo {pages} {2445--2459} (\bibinfo {year} {1974})}\BibitemShut {NoStop}%
\bibitem [{\citenamefont {Ercolessi}\ \emph {et~al.}(2018)\citenamefont
  {Ercolessi}, \citenamefont {Facchi}, \citenamefont {Magnifico}, \citenamefont
  {Pascazio},\ and\ \citenamefont {Pepe}}]{PhysRevD.98.074503}%
  \BibitemOpen
  \bibfield  {author} {\bibinfo {author} {\bibfnamefont {E.}~\bibnamefont
  {Ercolessi}}, \bibinfo {author} {\bibfnamefont {P.}~\bibnamefont {Facchi}},
  \bibinfo {author} {\bibfnamefont {G.}~\bibnamefont {Magnifico}}, \bibinfo
  {author} {\bibfnamefont {S.}~\bibnamefont {Pascazio}}, \ and\ \bibinfo
  {author} {\bibfnamefont {F.~V.}\ \bibnamefont {Pepe}},\ }\bibfield  {title}
  {\enquote {\bibinfo {title} {Phase transitions in ${Z}_{n}$ gauge models:
  Towards quantum simulations of the schwinger-weyl qed},}\ }\href {\doibase
  10.1103/PhysRevD.98.074503} {\bibfield  {journal} {\bibinfo  {journal} {Phys.
  Rev. D}\ }\textbf {\bibinfo {volume} {98}},\ \bibinfo {pages} {074503}
  (\bibinfo {year} {2018})}\BibitemShut {NoStop}%
\bibitem [{\citenamefont {Levin}\ and\ \citenamefont {Wen}(2005)}]{Levin2005}%
  \BibitemOpen
  \bibfield  {author} {\bibinfo {author} {\bibfnamefont {M.~A.}\ \bibnamefont
  {Levin}}\ and\ \bibinfo {author} {\bibfnamefont {X.~G.}\ \bibnamefont
  {Wen}},\ }\bibfield  {title} {\enquote {\bibinfo {title} {String-net
  condensation: A physical mechanism for topological phases},}\ }\href
  {\doibase 10.1103/PhysRevB.71.045110} {\bibfield  {journal} {\bibinfo
  {journal} {Phys. Rev. B}\ }\textbf {\bibinfo {volume} {71}},\ \bibinfo
  {pages} {045110} (\bibinfo {year} {2005})}\BibitemShut {NoStop}%
\bibitem [{\citenamefont {Wen}(2005)}]{Wen2005}%
  \BibitemOpen
  \bibfield  {author} {\bibinfo {author} {\bibfnamefont {X.~G.}\ \bibnamefont
  {Wen}},\ }\bibfield  {title} {\enquote {\bibinfo {title} {An introduction to
  quantum order, string-net condensation, and emergence of light and
  fermions},}\ }\href {\doibase https://doi.org/10.1016/j.aop.2004.07.001}
  {\bibfield  {journal} {\bibinfo  {journal} {Ann. Phys.}\ }\textbf {\bibinfo
  {volume} {316}},\ \bibinfo {pages} {1} (\bibinfo {year} {2005})}\BibitemShut
  {NoStop}%
\bibitem [{\citenamefont {Fradkin}(2013)}]{fradkin_2013}%
  \BibitemOpen
  \bibfield  {author} {\bibinfo {author} {\bibfnamefont {E.}~\bibnamefont
  {Fradkin}},\ }\href {\doibase 10.1017/CBO9781139015509} {\emph {\bibinfo
  {title} {Field Theories of Condensed Matter Physics}}},\ \bibinfo {edition}
  {2nd}\ ed.\ (\bibinfo  {publisher} {Cambridge University Press},\ \bibinfo
  {year} {2013})\BibitemShut {NoStop}%
\bibitem [{\citenamefont {Kitaev}(2003)}]{Kitaev_2003}%
  \BibitemOpen
  \bibfield  {author} {\bibinfo {author} {\bibfnamefont {A.}~\bibnamefont
  {Kitaev}},\ }\bibfield  {title} {\enquote {\bibinfo {title} {Fault-tolerant
  quantum computation by anyons},}\ }\href {\doibase
  https://doi.org/10.1016/S0003-4916(02)00018-0} {\bibfield  {journal}
  {\bibinfo  {journal} {Ann. Phys.}\ }\textbf {\bibinfo {volume} {303}},\
  \bibinfo {pages} {2 -- 30} (\bibinfo {year} {2003})}\BibitemShut {NoStop}%
\bibitem [{\citenamefont {Kitaev}\ and\ \citenamefont
  {Laumann}(2009)}]{Kitaev_2009}%
  \BibitemOpen
  \bibfield  {author} {\bibinfo {author} {\bibfnamefont {A.}~\bibnamefont
  {Kitaev}}\ and\ \bibinfo {author} {\bibfnamefont {C.}~\bibnamefont
  {Laumann}},\ }\bibfield  {title} {\enquote {\bibinfo {title} {Topological
  phases and quantum computation},}\ }\href
  {https://ui.adsabs.harvard.edu/abs/2009arXiv0904.2771K} {\bibfield  {journal}
  {\bibinfo  {journal} {arXiv e-prints}\ ,\ \bibinfo {pages} {arXiv:0904.2771}}
  (\bibinfo {year} {2009})}\BibitemShut {NoStop}%
\bibitem [{\citenamefont {Fowler}\ \emph {et~al.}(2012)\citenamefont {Fowler},
  \citenamefont {Mariantoni}, \citenamefont {Martinis},\ and\ \citenamefont
  {Cleland}}]{Scode_2012}%
  \BibitemOpen
  \bibfield  {author} {\bibinfo {author} {\bibfnamefont {A.~G.}\ \bibnamefont
  {Fowler}}, \bibinfo {author} {\bibfnamefont {M.}~\bibnamefont {Mariantoni}},
  \bibinfo {author} {\bibfnamefont {J.~M.}\ \bibnamefont {Martinis}}, \ and\
  \bibinfo {author} {\bibfnamefont {A.~N.}\ \bibnamefont {Cleland}},\
  }\bibfield  {title} {\enquote {\bibinfo {title} {Surface codes: Towards
  practical large-scale quantum computation},}\ }\href {<Go to
  ISI>://WOS:000308863900004
  https://journals.aps.org/pra/pdf/10.1103/PhysRevA.86.032324} {\bibfield
  {journal} {\bibinfo  {journal} {Phys. Rev. A}\ }\textbf {\bibinfo {volume}
  {86}} (\bibinfo {year} {2012})}\BibitemShut {NoStop}%
\bibitem [{\citenamefont {Zohar}\ \emph {et~al.}(2017)\citenamefont {Zohar},
  \citenamefont {Farace}, \citenamefont {Reznik},\ and\ \citenamefont
  {Cirac}}]{Zohar2017}%
  \BibitemOpen
  \bibfield  {author} {\bibinfo {author} {\bibfnamefont {E.}~\bibnamefont
  {Zohar}}, \bibinfo {author} {\bibfnamefont {A.}~\bibnamefont {Farace}},
  \bibinfo {author} {\bibfnamefont {B.}~\bibnamefont {Reznik}}, \ and\ \bibinfo
  {author} {\bibfnamefont {J.~I.}\ \bibnamefont {Cirac}},\ }\bibfield  {title}
  {\enquote {\bibinfo {title} {Digital quantum simulation of z(2) lattice gauge
  theories with dynamical fermionic matter},}\ }\href {\doibase
  10.1103/PhysRevLett.118.070501} {\bibfield  {journal} {\bibinfo  {journal}
  {Phys. Rev. Lett.}\ }\textbf {\bibinfo {volume} {118}},\ \bibinfo {pages} {5}
  (\bibinfo {year} {2017})}\BibitemShut {NoStop}%
\bibitem [{\citenamefont {Bender}\ \emph {et~al.}(2018)\citenamefont {Bender},
  \citenamefont {Zohar}, \citenamefont {Farace},\ and\ \citenamefont
  {Cirac}}]{Bender_2018}%
  \BibitemOpen
  \bibfield  {author} {\bibinfo {author} {\bibfnamefont {J.}~\bibnamefont
  {Bender}}, \bibinfo {author} {\bibfnamefont {E.}~\bibnamefont {Zohar}},
  \bibinfo {author} {\bibfnamefont {A.}~\bibnamefont {Farace}}, \ and\ \bibinfo
  {author} {\bibfnamefont {J~I.}\ \bibnamefont {Cirac}},\ }\bibfield  {title}
  {\enquote {\bibinfo {title} {Digital quantum simulation of lattice gauge
  theories in three spatial dimensions},}\ }\href {\doibase
  10.1088/1367-2630/aadb71} {\bibfield  {journal} {\bibinfo  {journal} {New J.
  Phys.}\ }\textbf {\bibinfo {volume} {20}},\ \bibinfo {pages} {093001}
  (\bibinfo {year} {2018})}\BibitemShut {NoStop}%
\bibitem [{\citenamefont {Lamm}\ \emph {et~al.}(2019)\citenamefont {Lamm},
  \citenamefont {Lawrence},\ and\ \citenamefont {Yamauchi}}]{Lamm2019}%
  \BibitemOpen
  \bibfield  {author} {\bibinfo {author} {\bibfnamefont {H.}~\bibnamefont
  {Lamm}}, \bibinfo {author} {\bibfnamefont {S.}~\bibnamefont {Lawrence}}, \
  and\ \bibinfo {author} {\bibfnamefont {Y.}~\bibnamefont {Yamauchi}} (\bibinfo
  {collaboration} {NuQS Collaboration}),\ }\bibfield  {title} {\enquote
  {\bibinfo {title} {General methods for digital quantum simulation of gauge
  theories},}\ }\href {\doibase 10.1103/PhysRevD.100.034518} {\bibfield
  {journal} {\bibinfo  {journal} {Phys. Rev. D}\ }\textbf {\bibinfo {volume}
  {100}},\ \bibinfo {pages} {034518} (\bibinfo {year} {2019})}\BibitemShut
  {NoStop}%
\bibitem [{\citenamefont {{Schweizer}}\ \emph {et~al.}(2019)\citenamefont
  {{Schweizer}}, \citenamefont {{Grusdt}}, \citenamefont {{Berngruber}},
  \citenamefont {{Barbiero}}, \citenamefont {{Demler}}, \citenamefont
  {{Goldman}}, \citenamefont {{Bloch}},\ and\ \citenamefont
  {{Aidelsburger}}}]{Schweizer_2019}%
  \BibitemOpen
  \bibfield  {author} {\bibinfo {author} {\bibfnamefont {C.}~\bibnamefont
  {{Schweizer}}}, \bibinfo {author} {\bibfnamefont {F.}~\bibnamefont
  {{Grusdt}}}, \bibinfo {author} {\bibfnamefont {M.}~\bibnamefont
  {{Berngruber}}}, \bibinfo {author} {\bibfnamefont {L.}~\bibnamefont
  {{Barbiero}}}, \bibinfo {author} {\bibfnamefont {E.}~\bibnamefont
  {{Demler}}}, \bibinfo {author} {\bibfnamefont {N.}~\bibnamefont {{Goldman}}},
  \bibinfo {author} {\bibfnamefont {I.}~\bibnamefont {{Bloch}}}, \ and\
  \bibinfo {author} {\bibfnamefont {M.}~\bibnamefont {{Aidelsburger}}},\
  }\bibfield  {title} {\enquote {\bibinfo {title} {{Floquet approach to
  $\mathbb{Z}_{2}$ lattice gauge theories with ultracold atoms in optical
  lattices}},}\ }\href {\doibase 10.1038/s41567-019-0649-7} {\bibfield
  {journal} {\bibinfo  {journal} {Nature Physics}\ }\textbf {\bibinfo {volume}
  {15}},\ \bibinfo {pages} {1168--1173} (\bibinfo {year} {2019})}\BibitemShut
  {NoStop}%
\bibitem [{\citenamefont {{Cui}}\ \emph {et~al.}(2020)\citenamefont {{Cui}},
  \citenamefont {{Yang}},\ and\ \citenamefont {{Shi}}}]{qz2_cui}%
  \BibitemOpen
  \bibfield  {author} {\bibinfo {author} {\bibfnamefont {X.}~\bibnamefont
  {{Cui}}}, \bibinfo {author} {\bibfnamefont {J.~C.}\ \bibnamefont {{Yang}}}, \
  and\ \bibinfo {author} {\bibfnamefont {Y.}~\bibnamefont {{Shi}}},\ }\bibfield
   {title} {\enquote {\bibinfo {title} {{Circuit-based digital adiabatic
  quantum simulation and pseudoquantum simulation as new approaches to lattice
  gauge theory}},}\ }\href@noop {} {\bibfield  {journal} {\bibinfo  {journal}
  {Journal of High Energy Physics}\ } (\bibinfo {year} {2020})}\BibitemShut
  {NoStop}%
\bibitem [{\citenamefont {Trotter}(1959)}]{Trotter1959}%
  \BibitemOpen
  \bibfield  {author} {\bibinfo {author} {\bibfnamefont {H.~F.}\ \bibnamefont
  {Trotter}},\ }\bibfield  {title} {\enquote {\bibinfo {title} {On the product
  of semi-groups of operators},}\ }\href {www.jstor.org/stable/2033649}
  {\bibfield  {journal} {\bibinfo  {journal} {Proc. Am. Math. Soc.}\ }\textbf
  {\bibinfo {volume} {10}},\ \bibinfo {pages} {545} (\bibinfo {year}
  {1959})}\BibitemShut {NoStop}%
\bibitem [{\citenamefont {Lloyd}(1996)}]{Lloyd1996}%
  \BibitemOpen
  \bibfield  {author} {\bibinfo {author} {\bibfnamefont {S.}~\bibnamefont
  {Lloyd}},\ }\bibfield  {title} {\enquote {\bibinfo {title} {Universal quantum
  simulators},}\ }\href {\doibase 10.1126/science.273.5278.1073} {\bibfield
  {journal} {\bibinfo  {journal} {Science}\ }\textbf {\bibinfo {volume}
  {273}},\ \bibinfo {pages} {1073} (\bibinfo {year} {1996})}\BibitemShut
  {NoStop}%
\bibitem [{\citenamefont {Hatano}\ and\ \citenamefont
  {Suzuki}(2005)}]{Hatano2005}%
  \BibitemOpen
  \bibfield  {author} {\bibinfo {author} {\bibfnamefont {N.}~\bibnamefont
  {Hatano}}\ and\ \bibinfo {author} {\bibfnamefont {M.}~\bibnamefont
  {Suzuki}},\ }\enquote {\bibinfo {title} {Finding exponential product formulas
  of higher orders},}\ in\ \href {\doibase 10.1007/11526216_2} {\emph {\bibinfo
  {booktitle} {Quantum Annealing and Other Optimization Methods}}}\ (\bibinfo
  {publisher} {Springer Berlin Heidelberg},\ \bibinfo {address} {Berlin,
  Heidelberg},\ \bibinfo {year} {2005})\ p.~\bibinfo {pages} {37}\BibitemShut
  {NoStop}%
\bibitem [{\citenamefont {{Childs}}\ \emph {et~al.}(2019)\citenamefont
  {{Childs}}, \citenamefont {{Su}}, \citenamefont {{Tran}}, \citenamefont
  {{Wiebe}},\ and\ \citenamefont {{Zhu}}}]{Childs_2019}%
  \BibitemOpen
  \bibfield  {author} {\bibinfo {author} {\bibfnamefont {A.~M.}\ \bibnamefont
  {{Childs}}}, \bibinfo {author} {\bibfnamefont {Y.}~\bibnamefont {{Su}}},
  \bibinfo {author} {\bibfnamefont {M.~C.}\ \bibnamefont {{Tran}}}, \bibinfo
  {author} {\bibfnamefont {N.}~\bibnamefont {{Wiebe}}}, \ and\ \bibinfo
  {author} {\bibfnamefont {S.}~\bibnamefont {{Zhu}}},\ }\bibfield  {title}
  {\enquote {\bibinfo {title} {{A Theory of Trotter Error}},}\ }\href@noop {}
  {\bibfield  {journal} {\bibinfo  {journal} {arXiv e-prints}\ ,\ \bibinfo
  {eid} {arXiv:1912.08854}} (\bibinfo {year} {2019})}\BibitemShut {NoStop}%
\bibitem [{\citenamefont {Shi}\ and\ \citenamefont {Wu}(2004)}]{shi}%
  \BibitemOpen
  \bibfield  {author} {\bibinfo {author} {\bibfnamefont {Y.}~\bibnamefont
  {Shi}}\ and\ \bibinfo {author} {\bibfnamefont {Y.~S.}\ \bibnamefont {Wu}},\
  }\bibfield  {title} {\enquote {\bibinfo {title} {Perturbative formulation and
  nonadiabatic corrections in adiabatic quantum-computing schemes},}\ }\href
  {https://link.aps.org/doi/10.1103/PhysRevA.69.024301} {\bibfield  {journal}
  {\bibinfo  {journal} {Phys. Rev. A}\ }\textbf {\bibinfo {volume} {69}},\
  \bibinfo {pages} {024301} (\bibinfo {year} {2004})}\BibitemShut {NoStop}%
\bibitem [{\citenamefont {{Jones}}\ \emph {et~al.}(2019)\citenamefont
  {{Jones}}, \citenamefont {{Brown}}, \citenamefont {{Bush}},\ and\
  \citenamefont {{Benjamin}}}]{quest}%
  \BibitemOpen
  \bibfield  {author} {\bibinfo {author} {\bibfnamefont {T.}~\bibnamefont
  {{Jones}}}, \bibinfo {author} {\bibfnamefont {A.}~\bibnamefont {{Brown}}},
  \bibinfo {author} {\bibfnamefont {I.}~\bibnamefont {{Bush}}}, \ and\ \bibinfo
  {author} {\bibfnamefont {S.~C.}\ \bibnamefont {{Benjamin}}},\ }\bibfield
  {title} {\enquote {\bibinfo {title} {{QuEST and High Performance Simulation
  of Quantum Computers}},}\ }\href {\doibase 10.1038/s41598-019-47174-9}
  {\bibfield  {journal} {\bibinfo  {journal} {Sci. Rep.}\ }\textbf {\bibinfo
  {volume} {9}},\ \bibinfo {eid} {10736} (\bibinfo {year} {2019})}\BibitemShut
  {NoStop}%
\end{thebibliography}%

\end{document}